\documentclass[final,5p,times]{elsarticle}

\usepackage{graphics}

\usepackage{amssymb}

\journal{Physica D}
\usepackage{color}

\begin{document}

\begin{frontmatter}



\title{Two-dimensional wave patterns of spreading depolarization:\\retracting, re-entrant, and stationary waves \tnoteref{label1}}

\tnotetext[label1]{Special issue: Emerging Phenomena.}

\author[ITP,KN2,IfN]{Markus A. Dahlem\corref{cor1}}
\author[MPI]{Rudolf Graf}
\author[STRONG]{Anthony J. Strong}
\author[CHARITE]{Jens P. Dreier}
\author[CHARITE,HOHENHEIM]{Yuliya A. Dahlem}
\author[HOHENHEIM]{Michaela Sieber}
\author[HOHENHEIM]{Wolfgang Hanke}
\author[AACHEN]{Klaus Podoll}
\author[ITP]{Eckehard Sch\"oll}

\cortext[cor1]{Corresponding author, dahlem@physik.tu-berlin.de}

\address[ITP]{Institut f{\"u}r Theoretische Physik, Technische
Universit{\"a}t Berlin,   Berlin, Germany}

\address[KN2]{Department of Neurology,
  Otto-von-Guericke-University Magdeburg, Magdeburg, Germany}
  \address[IfN]{Leibniz Institute f{\"u}r Neurobiologie, 39118 Magdeburg, Germany}

\address[MPI]{Max Planck Institute for Neurological Research, Cologne, Germany}

\address[STRONG]{King’s College London, Department of Clinical Neuroscience, Institute
of Psychiatry, United}

\address[CHARITE]{Department of Neurology, Charit\'e Campus Mitte, Berlin, Germany}

\address[HOHENHEIM]{University of Hohenheim, Membrane Physiology,
Stuttgart, Germany}

\address[AACHEN]{Department of Psychiatry and Psychotherapy, RWTH Aachen
University, Aachen, Germany}

\begin{abstract}
We present spatio-temporal characteristics of spreading depolarizations (SD) in
two experimental systems:  retracting SD wave segments observed with intrinsic
optical signals in chicken retina, and  spontaneously occurring re-entrant SD
waves that repeatedly spread across gyrencephalic feline cortex observed by
laser speckle flowmetry.  A mathematical framework of reaction-diffusion
systems with augmented transmission capabilities is developed to explain the
emergence  and transitions between these patterns.  Our prediction is that the
observed patterns are reaction-diffusion patterns controlled and modulated by
weak nonlocal coupling. The described spatio-temporal characteristics of SD are
of important clinical relevance under conditions of migraine and stroke.  In
stroke, the emergence of re-entrant SD waves is believed to worsen outcome.  In
migraine, retracting SD wave segments cause neurological symptoms and
transitions to stationary SD wave patterns may cause persistent symptoms without
evidence from  noninvasive imaging of infarction.  \end{abstract}


\end{frontmatter}


\section{Introduction}
\label{sec:introduction}

%
%
The World Health Organization lists migraines  among the four  most disabling
chronic medical disorders \cite{DOD08}.  Stroke is currently the third leading
cause of death in developed countries. There is  clinical and
experimental evidence that in both conditions spreading depolarization (SD)
waves of non-or-all type emerge.  SD waves are characterized by massive
redistribution of ions across cell membranes peaking after several seconds in a
nearly complete neuronal depolarization, followed by a much slower recovery
process taking up to minutes during which ion gradients are re-established
towards their physiological values. This maximal ionic perturbation clearly
distinguishes SD from all other brain states such as epileptic seizure
activity, functional activation or the physiological resting state. The
sequence of ionic perturbation and its recovery  spreads at a pace of about
3~mm/min over cortical regions.

In migraine, this phenomenon---originally termed {\em spreading depression}
of activity in the cerebral cortex \cite{LEA44},   causes
neurological symptoms starting about 30~min before the headache phase
\cite{MIL58,LAU87,HAD01}.  It is also currently debated whether migraine headache is caused
by substances released in the course of SD \cite{MOS07}. In stroke, so-called
{\em periinfarct depolarizations} \cite{HOS96}, a variant of SD, occur and spread in
the boundary zones surrounding a cerebral infarct.  A greater number of SD
waves is believed to worsen outcome.  SD has been demonstrated in patients
suffering from traumatic brain injury, ischemic stroke, subarachnoid
haemorrhage and intracerebral haematoma \cite{HOS96,STR02a,FAB06,DRE06,DOH08}.

The difference between SD in migraine and stroke is that SD in the latter occurs
when the cortex is deprived of adequate perfusion, i.\,e., under ischemia,
while SD in migraine occurs under norm\-oxic conditions, but otherwise both
processes share a common underlying neuronal mechanism \cite{SOM01}, and they
should not be seen as two different processes. For this reason they are
referred to as {\em spreading depolarizations}  \cite{DRE06}.  This name
reflects the common mechanism rather than the effect (depression of activity)
or location (periphery of infarct) of these waves as the other terms do.
To focus on common macroscopic features is in particular advisable when
two-dimensional spatio-temporal characteristics are investigated.  Moreover, it
emphasises the depolarization wave as one of the common aspects within the
complex bidirectional relation between migraine and stroke \cite{OLE93,MOS08}.
For example, SD  activity may play an important role in complications of
migraine, including migrainous infarction, persisting migraine aura without
infarction, and ischaemia-induced migraine (symptomatic migraine).

In this study, our aim is to understand the emergence of pathological
spatio-temporal SD states in the cortex and possible transitions between them
in terms of nonlinear dynamics and pattern formation. This approach complements
more traditional clinical research methodologies.  The paper is organized as
follows. In Sec.~\ref{sec:methods}, we introduce the experimental setup,
followed by a description of the observed patterns in
Sec.~\ref{sec:experimentalResults}. In Sec.~\ref{sec:mm}, we propose a
mathematical framework to model reaction-diffusion patterns that are controlled
and modulated by weak nonlocal transmission, and  in Sec.~\ref{sec:atCases} we
illustrate the applicability of this concept by special cases derived from this
framework. We end with a discussion in Sec.~\ref{sec:discussion}.

\section{Methods and Materials} \label{sec:methods}

Monitoring the spatio-temporal characteristics of SD  in two spatial dimensions
is crucial for studying this phenomenon, in particular from a clinical point of
view. We present data from two experimental systems of complementary complexity
each with a specific optical method.  Optical methods have much better spatial
resolution than electrical measurements. They provide indirect evidence of SD.
The information about the spatio-temporal development of SD is  very precise
since correlation between the optical signals and  electrophysiological changes of SD
are well studied \cite{PEI01,DUN01a,STR07}.


\subsection{Chicken retina as SD model} \label{sec:chicken}

The chicken retina is a typical {\it in vitro} system used to study the
formation of pathological activity pattern of SD by intrinsic optical signals
\cite{LIM99,DAH00b}. Retinal SD can be easily observed by intrinsic optical
signals, because it  evokes an optical signal with an amplitude being several
orders of magnitude higher than that observed in cortex. In fact, retinal SD
can be observed by the naked eye as a milky front. Moreover, the chicken retina
is a very homogeneous neural tissue because it lacks blood vessels. The neural
aspects of pattern formation can be thus studied in retinal SD in isolation,
that is, without the neurovascular coupling, with the obvious disadvantage that
retinal SD provides no information on this coupling.  Last but not least, the
retina has, similar to the cortex, a layered structure with several layers of
neurons interconnected by synapses and there is a solid knowledge of retinal network functioning  \cite{MIL08}.

The chickens (4-7 days) were decapitated and the eyes were carefully extracted
from the orbit. The eye was purged from adherent muscles and connective
tissue and afterwards the eye was equatorially divided. The vitreous body was
gently removed, and then the eyecup including the retina was transferred into a
small petri-dish filled with standard Ringer (R) solution
(Fig~\ref{fig:rSDStimMgBreak} (a)).  The standard Ringer solution
has the following composition (in $mM$): $100\;NaCl$, $6\;KCl$, $1\;CaCl_2$,
$1\;MgSO_4$, $30\;NaHCO_3$, $1\;NaH_2PO_4$,  $30$  glucose.  Throughout the 
experiment,  the submerged retina was perfused with fresh Ringer solution. For
more details see Ref.~\cite{LIM99}.

Following this preparation protocol, the healthy chicken retina has a
dark appearance.  Since blood vessels do not course across the retinal surface,
the only landmark, situated in the infero-temporal field, is the pecten oculi
(Fig~\ref{fig:rSDStimMgBreak} (b)),  a highly vascularized organ
with corrugated structure that projects from the optic disc into the vitreous
body of the avian eye.  The pecten is believed to nourish the retina.
Pectineal blood capillaries  deliver nutrients to the retina, therefore, there
are no retinal blood vessels. This fact, together with the intrinsic optical
signal, makes the avian retina an ideal experimental system  to study SD.

All measurements were obtained with a CCD video camera and stored with a
DVD-recorder. The retina was pricked with a fine Tungsten needle to elicit a
retinal SD wave. After measuring a control wave, the standard Ringer solution
was replaced with a Ringer solution where drugs had been added.  After a time
of 30 minutes a new wave was elicited. The experiments in chicken retina  were approved by the 
local ethics committee (Stuttgart and Magdeburg) and the Regierungspr\"asident of Stuttgart and Magdeburg.

\subsection{Stroke model} \label{sec:cat}

The choice of the cat as a species for studying SD has been governed by
disappointing experience with the use of rodent models of stroke: none of
the promising trials of neuroprotective agents in rodents, especially those
that block SD, has been reproduced in clinical trials of the same agent.
The cat cortex is closer than rodent cortex to that of humans in two important
respects.  First, it is gyrencephalic (folded), whereas the rodent cortex is
lissencephalic (smooth). The data obtained  establish beyond doubt that SD
spreads via sulci (clefts between gyral convexities), and never "jumps" between
gyral convexities; the information has allowed reliable interpretation of data
from monitoring the human brain, and indeed provided a justification for the
initial work (with invasive monitoring methods) seeking evidence of
depolarisations in the injured human brain.  Secondly, the ratio of astrocytes
to neurones in the cat brain is much closer to that of the human brain than it
is in the rodent brain. This is important since the small energy metabolite
pool in the brain, glycogen, is almost entirely located in astrocytes, again
making a model of occlusive stroke in cats a more reliable analogue of human
stroke than one in rodents. A further advantage of the cat brain is that the
anatomical arrangement of its blood supply closely resembles that of the human
brain.

In cat cortex, SD can be visualized by illuminating the cortex with laser light
and imaging the resulting speckle pattern. The laser speckle contrast imaging
can be used  as a proxy for SD, because cerebral blood flow changes are almost
invariably coupled to SD and so these changes can be studied as surrogate
markers to image and map the occurrence and spread of SD in real time
\cite{DUN01a}.  The imaging observation field was chosen so as to achieve
maximal spatial resolution in the area around the periphery of the initial
stimulation site.  Fig.~\ref{fig:reentrantPID} (a) illustrates the anatomy of the
cat cerebral cortex, where the gyri are arranged in concentric arcs, in the
order ectosylvian (EG), suprasylvian (SG), marginal (MG) gyri. 
The experiments in cats were approved by the local ethics committee and the Regierungspr\"asident of Cologne.

\section{Experimental results} \label{sec:experimentalResults} 

In the following two subsections, we present  the observed SD patterns in two
experimental systems. We then develop in Sec.~\ref{sec:mm} the conceptual
framework for modeling SD on a macroscopic level that allows us to understand
these experimentally observed spatio-temporal  patterns of SD  and possible
transitions between them in terms of nonlinear dynamics. In Sec.\
\ref{sec:atCases}, special cases are derived from this framework. 

\subsection{Retracting retinal SD waves segments}
\label{sec:retractingWavesExperiment}

An SD wave segment is a wave with a broken wave front and thus with two freely
moving open ends. The open ends of the wave front allow more complex
spatio-temporal dynamics,  in contrast to a wave with a closed wave front or a
wave where the open ends are attached to a non-excitable boundary.  Monitoring
the dynamics of the open ends can be used to probe the tissue excitability, as
we will first show experimentally and then develop the mathematical framework
in the next section.

Travelling SD wave segments have, to our knowledge, neither been  observed in
cortex nor in any other grey matter brain region of any species.  However,
indirect evidence, presented in a recent study combining migraine patients'
reports and high-field functional magnetic resonance imaging, suggests that
retracting SD wave segments cause migraine aura symptoms \cite{DAH08d}.  We
show here that these retracting waves can be observed in retinal SD by
controlling excitability in a regime weakly susceptible to SD.  This provides
the first direct evidence that the recently predicted retracting waves
exist in SD and that their emergence depends critically on the tissue
excitability.

An SD wave segment can be created  from an initially closed circular SD wave
front. The circular SD wave  is created by local superthreshold stimulation.
One way  to break the SD front is by manually inducing a transient excitation
block in the tissue in some confined location in front of the approaching wave.
Performing this procedure, the two open ends of the wave segment usually curl
in to form a double spiral,  whose wave front then repeatedly re-enters the
same retinal tissue.  In chicken retina, a spiral-shaped SD can rotate up to 20 times
before its center collides with the tissue boundary \cite{DAH97}.  Another way
to obtain wave segments is to reduce retinal susceptibility to SD to a degree
at which the wave front spontaneously breaks. In this case, the newly emerged open
ends retract, as we show, and the wave vanishes invading the tissue only once.

\begin{figure}[t] \center
  \includegraphics[width=\columnwidth]{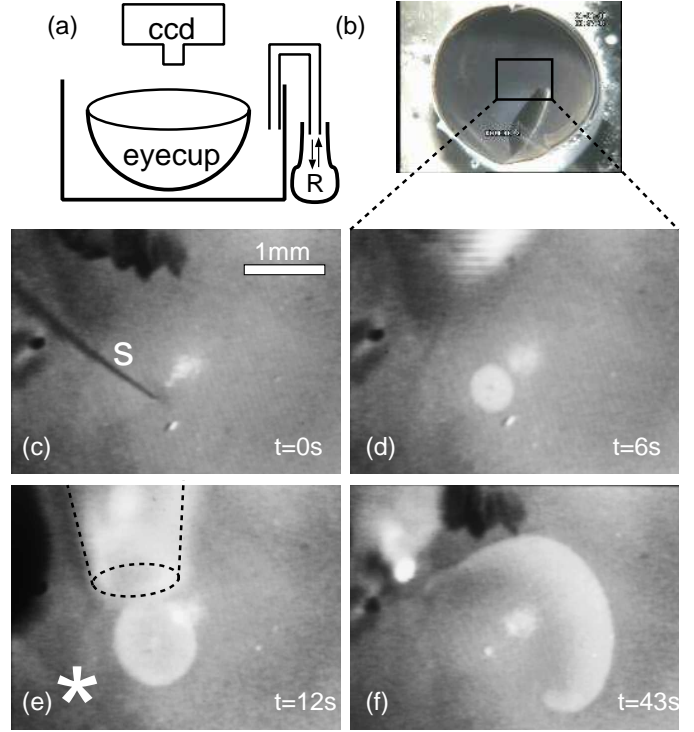}
  \caption{\label{fig:rSDStimMgBreak} Creation of a retinal SD wave
  segment with freely revolving  open ends: {\sf (a)} scheme of experimental
  set up with eyecup preparation placed in a petri-dish, perfused with
  Ringer (R) solution and monitored by a CCD camera; {\sf (b)} top view image
  from eyecup; {\sf (c)}  mechanical stimulation with sharp needle {\sf S};
  {\sf (d)} circular SD wave, {\sf (e)} local administration of $Mg^{2+}$ via
  pipette indicated by dashed lines at location marked by an asterisk; {\sf
  (f)} retinal SD wave segment.} \end{figure}

Firstly, we describe the manual creation of a spiral-shaped SD wave to contrast
this procedure and the resulting spatio-temporal develop\-ment with retracting
SD wave segments.  Retinal SD is elicited mechanically  by gently touching the
tissue with a sharp needle.  In Fig.~\ref{fig:rSDStimMgBreak} (c),
the stimulation needle (S) is seen by its dark shadow.  After about $15\,s$,
when  the extension of the circular SD wave front has grown in diameter to
about $0.75\,mm$, we administer  locally through a pipette a $0.5\,ml$ drop of
Ringer solution   containing a tenfold raised $Mg^{2+}$ concentration
($[Mg^{2+}]_R=10\,mM$). Pipettes having an inner tip diameter of about
$0.5\,mm$ are best suited.  They must be placed directly over the surface of
the submerged retina without touching it and less than $1\,mm$  in front of the
approaching SD wave (Fig.~\ref{fig:rSDStimMgBreak}~(e)).  For visual
guidance, the contours of the pipettes, which is out of focus in
Fig.~\ref{fig:rSDStimMgBreak}~(e), is  indicated by dashed lines and
its target position marked by an asterisk.  The propagation of the circular SD
wave is blocked around the pipette's target position and as a consequence its
wave front breaks open.  This wave segment develops into a spiral-shaped wave
(Fig.~\ref{fig:rSDStimMgBreak} (f)), if the submerged retina is kept
in standard Ringer solution and one end attaches to the tissue boundary of the
retina.

\begin{figure}[t] \center \includegraphics[width=0.95\columnwidth]{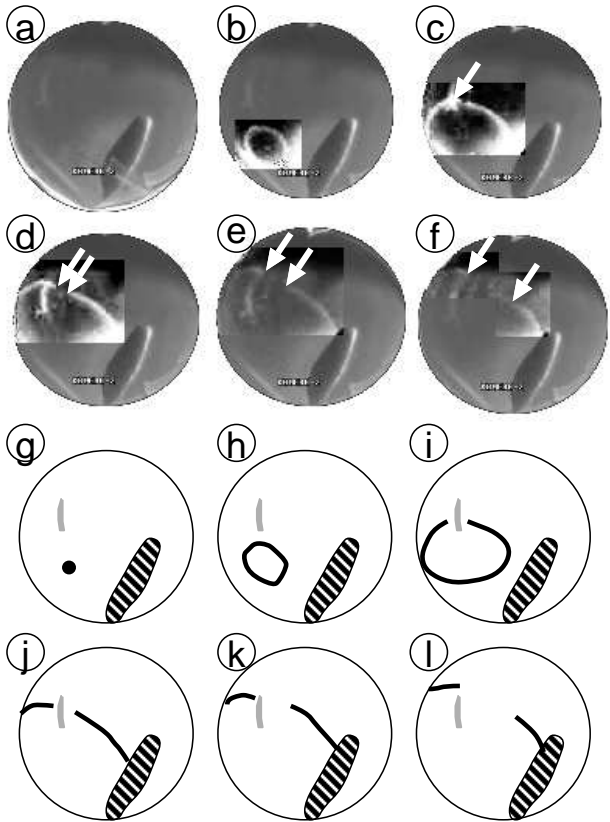}
  \caption{\label{fig:rSDCocaine} Six snapshots taken at $40\,s$
  time interval from chicken retina. (a)-(f) original recordings from CCD camera with enhanced
  brightness and contrast in  rectangular regions of interest. White arrow in
  (c) marks the breaking of the retinal SD wave, and subsequent pair of white
  arrows in (d)-(f) mark the location of the open ends of retinal SD wave.
  (g)-(l) Schematic illustration of this sequence. Retinal SD wave shown as
  black line, tissue in a compromised but viable state is marked gray, and
  pecten is shown filled with stripe pattern.} \end{figure}

We now adjusted retinal susceptibility to SD by adding cocaine in a
\mbox{$1-3\,mM$} concentration range in $1\,mM$ concentration steps. At $1\,mM$
cocaine concentration, SD waves propagated normally after being stimulated
mechanically. At $2\,mM$ cocaine concentration, we observed a spontaneous
excitation block.  In Fig.~\ref{fig:rSDCocaine} (c), this happens
at a location where the retina appearance was less dark indicating a
compromised tissue state.  The tissue at this location is, however, still
viable. After the cocaine is washed out, SD waves propagate normally again in
the entire retina.  The open ends of the SD wave segment retracts in the
further course (Fig.~\ref{fig:rSDCocaine} (d)-(f)).  In
Fig.~\ref{fig:rSDCocaine} (b)-(f), we manually and individually
enhanced the brightness and contrast in  rectangular regions of interest to
better visualize the SD wave.  Furthermore, in figures
(Fig.~\ref{fig:rSDCocaine} (g)-(l)) the spatio-temporal development
is schematically shown in a drawing.  

The retinal susceptibility to SD was at $2mM$ cocaine concentration already so
low, that in most cases the initiated wave {\em collapsed} after spreading only
a short distance. The term {\em collapse} refers to
an unstable, and therefore transient, SD wave profile that fades away although
the front does not break. This process was  predicted for SD to occur
extremely close to the susceptibility range of retracting waves segments
\cite{DAH07a}. At  $3mM$ cocaine concentration retinal SD waves could not be
elicited anymore.  Retracting retinal SD waves have also been observed by
reducing the potassium concentration in Ringer from $[K^{+}]_R=6\,mM$ to values
between $1-2\,mM$. For concentrations below $[K^{+}]_R=3\,mM$, the intrinsic
signal of the retinal SD wave profile changes in a characteristic way
\cite{DAH03a}, and the condition of the retina  was hardly stable for more than
30 min. Therefore no control measurement perfusing the retina with
standard Ringer could be done.

\begin{figure}[t] \center \includegraphics[width=0.8\columnwidth]{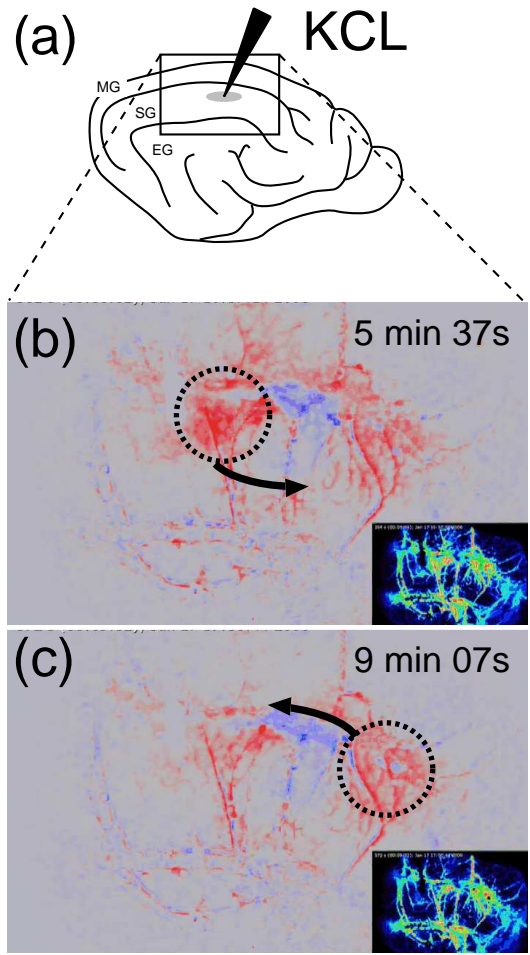}
  \caption{\label{fig:reentrantPID}  (a) Diagram of the lateral aspect of the right
  hemisphere of the cat brain illustrating typical gyral anatomy of a small ischaemic
  area (gray) within the area exposed for imaging (rectangle). (b) Following
  $KCL$ injection (at t=0), and  an immediate transient  radially spreading
  wave from the injection site,  a new single wave of hyperemia  appears
  spontaneously and cycles anticlockwise 3.5 times around an injured  area of some $3-5
  mm$ diameter centred on the original injection site, and commencing medial to
  it (above as viewed in this image field). Two snapshots were taken at a 3.5 min
  interval. The reader is advised to run the video to see the patterns.}
\end{figure}

\subsection{Re-entrant cortical SD waves}\label{sec:reentrantSD}

A single SD wave that repeatedly invades the same tissue  forms a re-entrant
pattern.  A re-entrant pattern is easy to understand as a wave propagating
around an obstacle. This obstacle can be a functional or anatomical block.  In
this subsection, we present  re-entrant SD patterns revolving around an
anatomical block that develops after microinjections of high concentration of
potassium have been locally administered to the cortical surface. Firstly, we
describe further in the next paragraph   functional blocks in SD  and
re-entrant spiral patterns, whose creation protocol was introduced  in the
previous subsection (Fig.~\ref{fig:rSDStimMgBreak} (f)). 

A functional block refers to a block that transiently forms as an intrinsic
property of the 2D wave dynamics. The core of the rotating spiral center in
retinal SD is an example of a functional block. Since re-entrant SD waves with
a functional block in the cortex have, to the best of our knowledge, not
been reported, we refer here to the observed characteristics in the retinal
experiments. The spiral core can be a circular area under certain conditions,
but it is in general a more complex structure. For example, the core does not
need to cover a spatially extended two-dimensional area, instead it can be an
one-dimensional curved line (inset (a) Fig.~\ref{fig:parameterPlane}), which
itself performs a rotation, as is the case for retinal SD \cite{DAH97}.  The
detailed dynamics of the core is a measure of susceptibility to SD and provides
important information about macroscopic model parameters (see
Sec.~\ref{sec:rd}).

In contrast to a functional block, an anatomical block refers to non-excitable
tissue, possibly an infarct region or an artificially created lesion. SD
patterns rotating around an artificially created lesion have also been called
reverberation of SD waves \cite{SHI72}. Reverberation of SD was demonstrated in
rat cerebral cortex and has been known for a long time. Recently, data from
monitoring the human brain has renewed interest in re-entrant patterns
\cite{DOH08}, in particular in experimental models of stroke to investigate the
cause of infarct  expansion, because more frequent  SD waves are believed
to worsen stroke outcome and  therefore the emergence of re-entrant SD waves is
of considerable clinical interest.

The creation of  a re-entrant wave  demands, in general, special initial
conditions.  SD is induced with microinjections into the cortex of $1\,M$
potassium chloride ($KCl$). As described in Sec.~\ref{sec:cat}, laser
speckle contrast imaging flowmetry  is used to detect waves of regional blood
flow alteration coupled to SD in a cortical field of view covering three gyri
called marginal, suprasylvian, and ectosylvian gyrus (MG, SG, and EG)
respectively (Fig.~\ref{fig:reentrantPID} (a)).  A radial pattern is regularly
observed for the initial wave after an injection, starting from the point of
$KCl$ injection and propagating outwards, often appearing as a concentric wave.
This is demonstrated with a supplementary
video (Video~1, filename dahlemEtAlVideo.mov). 

Following the radial pattern, an area of markedly reduced perfusion becomes
established surrounding the injection point. We infer that an ischemic region
develops at this site, most likely due to the marked vasoconstriction
associated with high levels of $K^+$ concentration in the perivascular fluid.
It is known that the perivascular administration of  $K^+$ concentrations at
levels of $40~mM$ decreases  pial arteries calibre by 47.2\% and vein calibre
by 13.5\% \cite{MCC82}.  This significant reduction changes the  conditions at
the injection point and in its border zone. At least this assumption is
consistent with the changes following the radial pattern and being observed in
the surrounding of this developing ischemic focus.  


During observation for 40 minutes following a $KCl$ injection, we saw  new
waves of hyperemia that developed spontaneously at the edge of the lesion,
without any further $KCl$ injection. Surprisingly, these secondary waves did
not propagate radially but instead each revolves several times the ischemic
focus.  This is demonstrated in the  supplementary video (Video~1, filename
dahlemEtAlVideo.mov). Similar perilesion re-entrant patterns have been seen around
experimental ischaemic lesions in the cerebral cortex after occlusion
\cite{STR07}. Furthermore, note that re-entrant patterns result in consistent
periodicity of depolarizations when monitored at a single cortical point, and
such periodic events have also been seen in the injured human brain
\cite{DOH08}.

\section{Augmented reaction-diffusion model}\label{sec:mm}

In this section, we develop the conceptual framework for modeling SD on a
macroscopic level.  This framework consists of a continuum model of the cortex
in which the communication between neurons is separated in two major schemes:
(i) local coupling, and (ii) nonlocal coupling in space and time. In the
cortex, these schemes represent (i) diffusion, and (ii) network communication
and neurovascular feedback.  With this framework  general information in gained
 on how reaction-diffusion patterns of phathological states in the cortex
are controlled and modulated by weak nonlocal transmission schemes.  We  apply
this framework in Sec.\ \ref{sec:atCases} to interpret the experimentally
observed patterns, in particular, their emergence and transitions.

\subsection{Model equation}

The level of detail and complexity of our framework is  determined  by both
spatial extension and propagation speed of the emerging patterns.  SD waves
extend in the cortex over several centimeters and spread with a remarkably slow
speed of several millimeters per minute.  The  extension of SD patterns
suggests that its mathematical description should be in terms of large-scale
activity in neural populations and describe how this activity affects the
cortical ionic equilibrium (homeostasis).  The slow speed of SD (about
$3\,mm\,min^{-1}= 50\,\mu m\,s^{-1}$) indicates that diffusion in cortical
tissue plays a major role.  Therefore, the central parts of a  macroscopic
continuum model of SD are reaction rate terms and diffusion terms. These terms
are supplemented by additional terms reflecting augmented transmission
capabilities.   Within such a framework, the model equation is
\begin{eqnarray}
\label{eq:main}
\left(\!
\begin{array}{*{1}{c}}
\partial_t u\\
\varepsilon^{-1}\partial_t v
\end{array}
\!\right)=
\left(\!
\begin{array}{*{1}{c}}
 f(u,v)\\
g(u,v)
\end{array}
\!\right) + 
{\bf D}\left(\!
\begin{array}{*{1}{c}}
u\\
v
\end{array}
\!\right)+ 
 \,\bf{H}\left(\!
\begin{array}{*{1}{c}}
u\\
v
\end{array}
\!\right),
\end{eqnarray}
where $\partial_t$ is a short hand notation for the derivative with respect
to time.  The terms on the right hand side represent (i) reaction rates
$f(u,v)$ and $g(u,v)$ of a simplified activator($u$)-inhibitor($v$) model of
cortical homeostasis (Sec.~\ref{sec:ai}), (ii) diffusion of chemical signals
and other local transmission schemes summarized under the term  {\em volume transmission}
in the brain and described by the diffusion operator $\bf D$
(Sec.~\ref{sec:vt}), and (iii) the augmented transmission capabilities
described by an operator $\bf H$ (Sec.\ \ref{sec:at}).  The model
equation~(\ref{eq:main}) and also its  derived special cases (Sec.\
\ref{sec:atCases}) are nondimensionalized equations. In such systems, natural
units of a system are used  in a systematic manner so that the input and output
variables of the  rate functions $f(u,v)$ and $g(u,v)$ are in the order of
unity. Therefore, beside many other advantages, nondimensionalization suggests
explicitly  a parameter, namely $\varepsilon$ that should be used for analyzing
this pattern forming system. The parameter $\varepsilon$ indicates how much
slower the production rate of the inhibitor $v$ is compared to the one of the
activator $u$.

%

\subsection{Activator-inhibitor dynamics}\label{sec:ai}

\begin{figure}[t] \center
  \includegraphics[width=0.95\columnwidth]{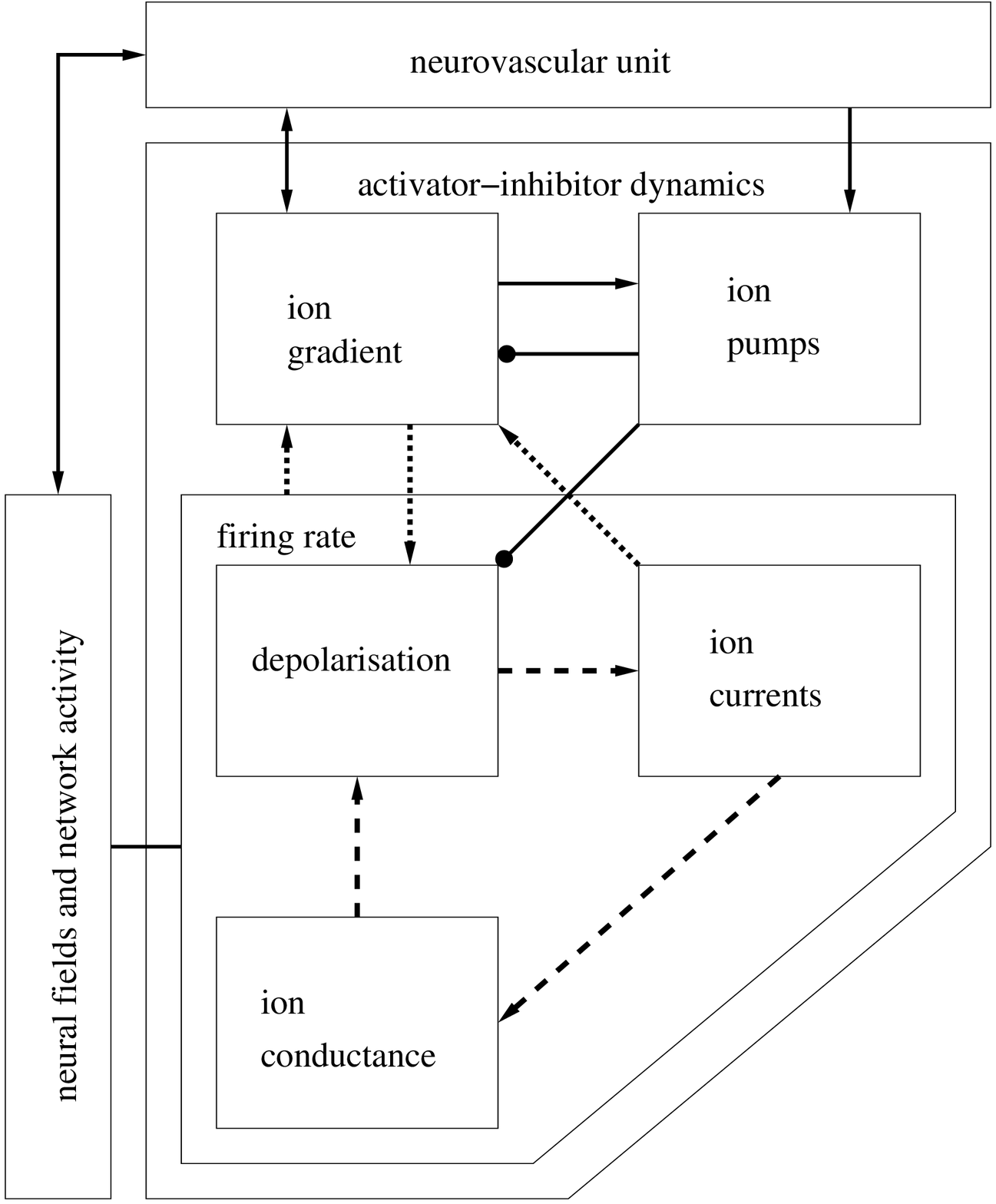}
  \caption{\label{fig:sdPathways} Flow chart of feedback loops and transmission pathways in SD. } \end{figure}

Local dynamics of  activator-inhibitor type are common in pattern forming
systems, and---if they are coupled by diffusion--have been shown to govern
several aspects of SD wave propagation \cite{GRA63,TUC81,REG94,REG96,DAH04b}.  The reaction rates
$f(u,v)$ and $g(u,v)$ in Eq.~(\ref{eq:main}) represent such simplified local dynamics
in the cortical ion homeostases (see Fig.~\ref{fig:sdPathways}) leading to
non-or-all type behavior (excitability).  The rate function $f(u,v)$ describes
the change in  a proxy  agent named activator $u$ by lumping together nonlinear
reactions containing a positive feedback loop in the SD mechanism, such as
$Na^+$-inward currents (dashed arrows in Fig.~\ref{fig:sdPathways}) and
extracellular potassium concentration $[K^+]_e$ (dotted arrows).  Likewise, the
rate function $g(u,v)$ of the inhibitor $v$ represents recovery processes,
such as effective regulation of $[K^+]_e$ and pump currents by the neuron's
electrogenic $Na^+$-$K^+$ ion pump (solid arrows with circular heads). 


An activator-inhibitor system exhibiting excitability requires a certain
configuration of trajectories that follow the vector field given by  the
functions  $f(u,v)$ and $g(u,v)$ in the phase space \cite{IZH00,SCH09c}.  This
configuration usually results from the parameter vicinity of an oscillatory
regime whose large amplitude limit cycle is suddenly destroyed.  A rest state
(fixed point) is usually the only attractor in the excitable regime.  There
exists also a threshold close to this rest state, e.\,g.,  a separatrix or a
sharp trajectory, that must be crossed by a stimulation to result in phase
space excursion. Activator-inhibitor systems are the most generic systems to
exhibit such behavior. A particular example is derived in Sec.\ \ref{sec:EHG}
and the validity and applicability of  generic  dynamics to model SD is
discussed in Sec.~\ref{sec:discussion}.

\subsection{Volume transmission}\label{sec:vt}

Communication between neurons is usually classified in two schemes: wiring and
volume transmission \cite{AGN95}.  Volume transmission is mainly characterized
by diffusion of chemical signals in the extracellular space.   That diffusion
is the main cause of the spread in SD is indicated by the  propagation speed of
about $50\,\mu m\,s^{-1}$. Waves caused by nonlinear reactions coupled by
diffusion of monovalent ions travel at speeds  in the order of $10\,\mu m
s^{-1}$ assuming moderate reaction rates in the nonlinear dynamics
\cite{DAH04b}.  Signals transmitted by cortical wiring, in particular  action
potential along axons, travel at speeds in the order of meters per second,
i.\,e., five orders of magnitude faster.  


The diffusion operator in Eq.~(\ref{eq:main}) and takes the general form
\begin{eqnarray}\label{eq:D} {\bf D}=\left(\begin{array}{*{2}{c}} 1 & 0\\ 0 &
  \lambda  \end{array} \right) \nabla^2, \end{eqnarray} with $\nabla^2$ being
  the Laplace operator.  We do not consider cross-diffusion, therefore
  off-diagonal entries vanish.  The foam-like geometry of  the interstitial space
  renders diffusion in the brain different from aqueous solution. In
  particular, apparent diffusion coefficients should be considered in a
  dimensionalized model \cite{SYK08}.  In our nondimensionalized framework, 
  diffusion coefficients of activator and inhibitor, $D_u$ and $D_v$,
  respectively, do not appear explicitly. Only the ratio
  $\lambda=\frac{D_v}{\varepsilon D_u}$ of the diffusion coefficients enters
  Eq.~(\ref{eq:D}) as a parameter (note that $\varepsilon$ appears also in this
  quotient).  In SD,  inhibitory reactions are thought to relate to ion pump
  activity (Fig.~\ref{fig:sdPathways}), in which case the inhibitor would not
  be subject to diffusion and therefore we choose $\lambda=0$ in this study.

\subsection{Augmented transmission}
\label{sec:at}

Similar to the diffusion operator in  Eq.~(\ref{eq:D}), the operator  $\bf H$
describes the augmented transmission capabilities as \begin{eqnarray}
  \label{eq:H} {\mathbf{H}}= K \left( \begin{array}{*{2}{c}}  A_{uu} & A_{uv}
    \\ A_{vu}  & A_{vv}  \end{array} \right) F. \end{eqnarray} 
$F$ is a operator specifying the augmented transmission (in analogy to the
Laplace operator $\nabla^2$ for volume transmission),
examples are given in Sec.\ \ref{sec:atCases}.  The four elements of the matrix
$A_{ij}$    specify the relative coupling strength among different schemes of
augmented transmission (in analogy to $\lambda$ in Eq.\ \ref{eq:D}) so that the operator $A_{ij}F$
represents three individual steps of augmented transmission, namely (I)
selecting a species $j$ whose transmission capability is augmented, (II)
creating the nonlocal driving force  from this species, and (III) feeding this
driving force back, multiplied by the coupling strength $K$, into the dynamical
variable of species $i$, with $i,j \in \{u,v\}$.

The parameter $K$ describes the coupling strength relative to the diffusion
coefficient, which is normalized to unity.  Therefore, the $A_{ij}$ are also
normalized to unity, i.\,e., 
\begin{equation} \label{eq:ANorm} \max(\{|A_{uu}|,
  |A_{uv}|, |A_{vu}|, |A_{vv}|\})=1 
\end{equation} 
and we limit our framework  to cases where 
\begin{equation} \label{eq:KLimit} 
  |K|\!<\!1 
\end{equation}
holds. With the constraints in Eqs.\ (\ref{eq:ANorm})-(\ref{eq:KLimit})
diffusion becomes the dominant coupling.  We pose these constraints, because,
within this framework, augmented transmission capabilities must not become an
essential part of the SD wave propagation mechanism. Rather we assume (i) that
the propagation of SD can be explained even for $K\!=\!0$, but (ii) that only
for $K\!\neq\!0$ and appropriate choices of $A_{ij}$ and $F$ the observed
two-dimensional SD patterns  are described correctly,
because augmented transmission determines the specific spatial
evolution of an SD wave on the two-dimensional cortical surface.

We pose one further constraint on augmented transmission, in addition to
$|A_{ij}| \leq 1$ and $|K|<1$, namely on the choices of $F$.   The operators
$F$ should be, in contrast to volume transmission, of nonlocal nature.  These
two constraints, summarized as {\em weak nonlocal coupling}, are the key to
incorporating  just the right amount of flexibility in this framework.  The
nonlocality,  on one hand,  addresses wiring transmission  by describing the
effect of the structural and functional cortical connectivity on SD, and, on
the other hand, includes  changes in cerebral blood flow in response to SD and
how these changes are fed back to change subsequent neural activity, a feedback
process termed neurovascular coupling.

Wiring transmission and neurovascular coupling can include nonlocalities in
both space and time. For wiring transmission,  nonlocalities in space are due
to long-range lateral cortical connections, and nonlocalities in time are due
to increased open probabilities of metabotropic receptors in the range of
seconds after their activation. The traveling speed of action potentials along
lateral cortical connections and the ionotropic synaptic transmission are,
compared to the speed of SD, quasi instantaneous.  Nonlocality in space for
neurovascular coupling, that is, how localized the vascular response is, also
referred to as the vascular point spread function,  depends on which component
of the vascular system is considered.  While changes to capillaries, i.\,e.,
the smallest component controlled by pericytes \cite{PEP06}, can be considered
as a localized transmission in space, compared to the SD wave size, changes to
arterioles and arteries have wider point spread functions.  Latencies in the
neurovascular coupling in the order of seconds are typical and cause
nonlocalities in time.  Moreover, wiring transmission and neurovascular
coupling can act in combination to constitute a nonlocal transmission scheme.
For example, the SD wave will activate quasi instantaneously a wide spread
increase in neural activity through feed-forward and feedback cortical
circuitry that induces therefore a spatially  global but time-delayed neurovascular response. 

In Sec.~\ref{sec:EHG} we introduce a first special case  derived from
Eq.~(\ref{eq:main}).  This is based on the Hodgkin-Grafstein model of SD 
extended by inhibitor dynamics. This results in a generic model with only
reaction-diffusion terms (i)-(ii). In Sec.~\ref{sec:rd}, we consider in this
generic system  emerging  patterns and transitions between them, in particular,
transitions from spiral waves to retracting wave segments.  In
Sec.~\ref{sec:atCases} we introduce  further cases  derived from
Eq.~(\ref{eq:main}) including a nonlocal term (iii).

\subsection{Extended Hodgkin-Grafstein model}\label{sec:EHG}

The first reaction-diffusion model of SD, suggested by Hodg\-kin and Grafstein
\cite{GRA63}, was a single species model. It is a special case model that can
be derived from our framework for $\varepsilon=0$. Then it follows that rate of
change $\partial_t v=0$ and thus the inhibitor $v$ is simply a constant
parameter.  The activator agent was suggested to be extracellular potassium
$[K^+]_e$.  We continue to use $u$ as the activator. Moreover, we want to
note that this activator can also be a combination of quantities having a
positive feedback loop, as shown in Fig.~\ref{fig:sdPathways}, without
restricting the main idea of the Hodgkin-Grafstein model.  The only assumption
is that the activator rate equation $f(u,v)$ is a cubic function in $u$  
\begin{eqnarray} 
  \label{eq:singleSpecies} 
  \frac{\partial u}{\partial t} &=&  u - \frac{u^3}{3}\; - v \;+\;\nabla^2 u.
\end{eqnarray} 
This model describes a bistable state in cortical ion homeostases. The smallest
and the largest roots of the cubic function are stable fixed points, the
lower one being identified as  the physiological state, the other one as the
pathological state. Furthermore,  the middle root  is an unstable fixed point,
i.\,e., a threshold. The Hogkin-Grafstein model equation
(\ref{eq:singleSpecies}) can be solved analytically and this provides detailed
mathematical insight in the mechanism of how the pathological state invades the
physiological state (see for example the textbook \cite{WIL99}).  

We can choose suitable inhibitor dynamics from the Fitz\-Hugh-Nagumo equations \cite{FIT61,NAG62} to extend the Hodgkin-Graf\-stein
model, i.\,e., to obtain a model with  $\varepsilon\!\neq\!0$. This way we
couple Eq.~(\ref{eq:singleSpecies}) to another ordinary differential equation
\begin{eqnarray} \label{eq:inhibitor_0} \varepsilon^{-1} \frac{\partial
  v}{\partial t}&=& u + \beta - \gamma v,  \end{eqnarray} 
where $\beta$ and $\gamma$  are two parameters. These parameter determine
the threshold, which in the single species model
(Eq.~(\ref{eq:singleSpecies})) was determined by $v$.  Usually, $\gamma$ is set
constant and then $\beta$ is solely identified as a threshold parameter. 

The FitzHugh-Nagumo equations  (\ref{eq:singleSpecies})-(\ref{eq:inhibitor_0})
are  a paradigmatic model for excitable media. An excitable medium has the
capacity to propagate a sustained wave as $u$ and $v$ concentration profiles
with a defined wave front and trailing edge through a spatially extended
system. Whereas the Hodgkin-Grafstein model can just account for front
propagation but not for  recovery defining the dynamics in the trailing edge.  

The biophysical interpretation of the variables and terms in the
FitzHugh-Nagumo equations ~(\ref{eq:singleSpecies})-(\ref{eq:inhibitor_0}) is
originally based on membrane dynamics in a single neuron \cite{FIT61,NAG62},
because these equations were derived from the Hodgkin-Huxley model of action
potentials.  The interpretation of
Eqs.~(\ref{eq:singleSpecies})-(\ref{eq:inhibitor_0}) as a model for SD is thus
very different.  In order not to  cause confusion, we call
Eqs.~(\ref{eq:singleSpecies})-(\ref{eq:inhibitor_0}) an extended
Hodgkin-Grafstein model in this study, although a great body of knowledge about
this generic system is published under the name   FitzHugh-Nagumo model (or
also under the name Bonhoeffer-van der Pol model \citep{POL26,POL29,BON48}).

The interpretation of the  terms on the right hand sides of
Eqs.~(\ref{eq:singleSpecies})-(\ref{eq:inhibitor_0}) can be directly read from
Fig.~\ref{fig:sdPathways}. In Eq.~(\ref{eq:singleSpecies}) the first "$+u$" on
the right hand side provides the positive self-feedback loop (dashed and dotted
loops), whereas the nonlinear cubic term 	restricts this autocatalytic process
and provides a threshold. The "$-v$" comes from the inhibitory effect of the
ion pumps (arrow with circular head between boxes marked as ion pump and ion
concentration). The main other important terms (as $\gamma$ can be set to a
constant in the range $0 \leq \gamma \leq 1$ without changing the generic
behavior) are the "$+u$" and the $\beta$ on the right hand side of
Eq.~(\ref{eq:inhibitor_0}). The $"+u"$  describes  positive pathway to the ion
pump activity when the  ion concentration is out of balance (solid arrow with pointed
head starting from box marked ion gradient and terminating at box ion pump), and the parameter
$\beta$ sets, as already mentioned, the threshold.

\subsection{Excitability probed by dynamics of functional block} 
\label{sec:rd}

\begin{figure}[t] \center
  \includegraphics[width=\columnwidth]{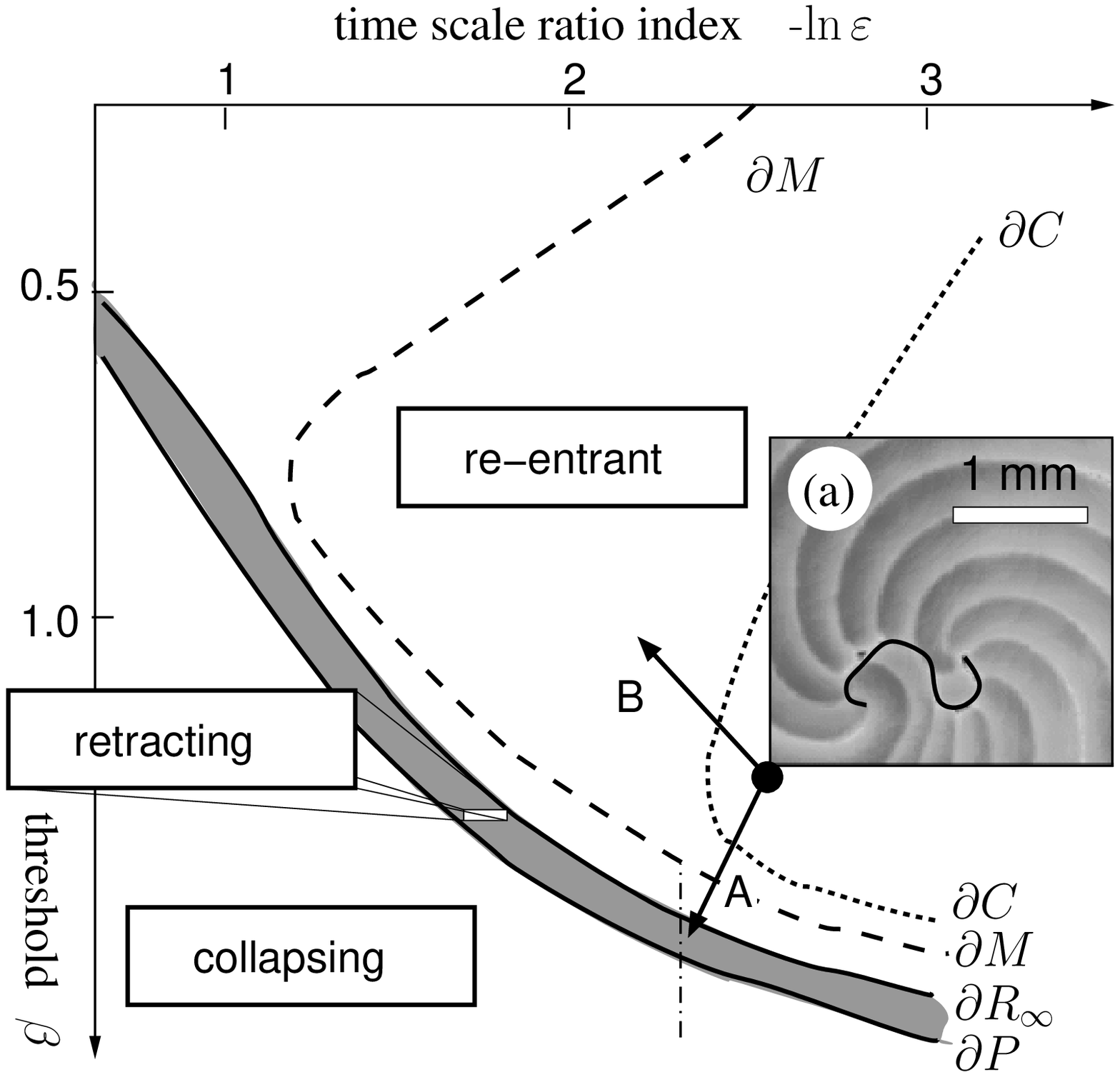}
  \caption{\label{fig:parameterPlane}  Parameter space of an generic active
  medium (FitzHugh-Nagumo equations adopted from \cite{WIN91}), with parameter
  regions separated by bifurcation lines  $\partial C$, $\partial M$, $\partial
  R_\infty$, and $\partial P$ that mark sudden changes in the spatio-temporal patterns
  from complex, meandering, rigid rotating spiral patterns, to the left of
  $\partial C$, $\partial M$ and $\partial R_\infty$, respectively, to the propagation
  boundary $\partial P$.  To the left of $\partial P$ the medium is
  non-excitable.   The inset (a) is the predicted location of retinal
  spiral-shaped SD waves  with a linear functional block (solid line) in the
  regime of complex re-entrant patterns. } \end{figure}

The extended Hodgkin-Grafstein model of SD  as a generic reaction-diffusion
model of activator-inhibitor type can explain the emergence and transition
between different spatio-temporal wave patterns in two spatial dimensions, for
example from re-entrant SD waves revolving around a functional block  to
retracting waves segments, by changing parameters in Eqs.\
(\ref{eq:singleSpecies})-(\ref{eq:inhibitor_0}).  This in turn can be
understood as a translation in the parameter space of this model.  If we set
$\gamma$ to a constant value  in the range $0 \leq \gamma \leq 1$, there
remain only two main parameters in this set of equations, namely the time scale
ratio $\varepsilon$ of inhibitor and activator dynamics and the threshold
$\beta$.  Both parameters are closely tied to the excitable behavior, that
emerges in active media.  Therefore, the parameter plane $(\varepsilon,\beta)$
can be viewed as the {\em excitability plane}, see
Fig.~\ref{fig:parameterPlane}, in which certain bifurcation lines mark the
emergence of specific spatio-temporal patterns.

One axis of the excitability plane is spanned by the parameter $\varepsilon$.
This parameter appears in Eq.~(\ref{eq:inhibitor_0}). It is also explicitly
given in the model equation~(\ref{eq:main}), indicating that $\varepsilon$ is
independent on the specific model derived from the framework. It is usually a
small positive number (\mbox{$0\!<\!\varepsilon\!\ll\!1$)}, which means that
the recovery processes (arrows with circular head in Fig.~\ref{fig:sdPathways})
are slower than the autocatalytic excitation caused by positive feedback loops
(dashed and dotted arrows in Fig.~\ref{fig:sdPathways}).  In
Fig.~\ref{fig:parameterPlane}, the horizontal axis of the excitability plane is
spanned by the negative logarithm of the time scale ratio $\varepsilon$, called
time scale ratio index $-\ln \varepsilon$. To give an example, if
$\varepsilon=0.1$, the kinetic rates of the inhibitory processes are  by  one
order of magnitude slower than those of the activator and the time scale
ratio index  is accordingly 1.
The other parameter, that spans the vertical axis of the excitability plane in
Fig.~\ref{fig:parameterPlane}, is the excitation threshold, which depends on
the specific model derived from the framework. In the extended
Hodgkin-Grafstein model   $\beta$ can be chosen as a threshold parameter, if
$\gamma$ is set constant, for example for Fig.~\ref{fig:parameterPlane} we
choose $\gamma=0.5$.  The units of $\beta$ are rather arbitrary, but a
characteristic value is $\beta=0.\bar 6$ (if $\gamma\!=\!0.5$) at which the
activator-inhibitor dynamics becomes oscillatory for small $\varepsilon$.

In
the parameter plane in Fig.\ \ref{fig:parameterPlane} certain bifurcation lines
mark the parameter values at which  specific  patterns emerge. These pattern are also seen in the
experimental SD systems by monitoring the functional block (spiral tip dynamics).
In fact, the dynamics of the functional block can be used to probe the
tissue excitability and locate the occurrence of these patterns in the excitability plane in
Fig.~\ref{fig:parameterPlane}. For example, it was shown that under
standard experimental condition (Sec.~\ref{sec:chicken}) the functional block
of retinal SD waves  is an one-dimensional curved line (inset (a)
Fig.~\ref{fig:parameterPlane}), which itself performs a complex rotation
\cite{DAH97}.  This is caused by a combination of effects. Essentially, a long
trailing edge (indicating small $\varepsilon$) causes a large refractory zone
that the spiral tip, which has a high tendency to curl in (indicating small
threshold $\beta$), must avoid leading to complex re-entrant manoeuvres.  Similar dynamics
of  linear functional blocks is obtained in computer simulations of the
extended Hodgkin-Grafstein model (i.\,e., Fitz\-Hugh-Nagumo equations
(\ref{eq:singleSpecies})-(\ref{eq:inhibitor_0})) in a certain parameter regime
close to the black full circle in Fig.~\ref{fig:parameterPlane} \cite{WIN91}.
This regime  is bounded from one side by a bifurcation line $\partial C$
(dotted line) beyond which complex rotation of largely  linear functional
blocks is found (to the right of $\partial C$).  On the other side of $\partial
C$, simpler interactions between spiral tip  and refractory zone occur and the
functional block is described by a meandering pattern. This region is bounded
by $\partial M$ (dashed line). Between yet another bifurcation line, $\partial
R_{\infty}$ (solid), and $\partial M$ occur reentrant patterns with a perfect circular
functional block (also called rigid rotation), indicating only weak or no interactions of the re-entrant
wave with its refractory  trailing edge. At least, in the narrow zone $\partial
R_\infty$ and $\partial P$ (shaded gray) only retracting wave segments occur.

\section{Control and modulation by weak nonlocal coupling}
\label{sec:atCases}

Control and modulation of reaction-diffusion patterns by weak nonlocal coupling
are the key features of the framework that provide enough flexibility to
explain the experimentally observed data.  In this section, we sketch principal
mechanisms of nonlocal transmission that lead to the driving force and are
hitherto hidden behind the nonlocal operator described by the terms
$A_{ij}\,F$. This augmented transmission can explain two-dimensional wave
patterns observed experimentally.  The emergence of these patterns depends on
this augmented transmission capabilities and would in this form not occur in a
pure reaction-diffusion system of activator-inhibitor type. Many of these
principles have been investigated in other active media in detail in previous
studies both experimentally and theoretically. For example, two-dimensional
pattern formation in the chemical Belousov-Zhabotinskii reaction was studied
with external feedback loops.  Therefore we only outline these principles and
refer to the corresponding studies for details.  We begin with describing
changes in excitability as  translation in the excitability plane, as shown in
Fig.~\ref{fig:parameterPlane}, along which the excitable state can  cross
bifurcation lines leading to an abrupt change in the two-dimensional wave
pattern (Sec.~\ref{sec:translation}).  Then, we consider the stabilization of
wave segments to account for sustained propagation
(Sec.~\ref{sec:stabilization}), and, finally investigate re-entrant wave
patterns in this framework  with further reference to clinical data of
persisting migraine aura without infarction (Sec.~\ref{sec:anatomical}). 



\subsection{Control by long-range connections and time-delayed feedback}
\label{sec:translation}

In this subsection, we describe the effect of long-range connections and
time-delayed feedback  on reaction-diffusion system of activator-inhibitor
type. This was reported in detail in recent studies \cite{DAH08,SCH09c}.
Therefore we can limit our description  to  issues that are relevant to link
this theoretical work to the experiments of SD waves in retina
(Sec.~\ref{sec:retractingWavesExperiment}),  in particular, to the transition
from spiral-shaped SD waves to SD wave segments by administering drugs (see
also Sec.  \ref{sec:rd}). In the outset of this section, we consider the
efficiency of general control schemes causing translations within the
excitability plane.

Let us assume, for the sake of argument, that we can deliberately control the
values of the parameters $\varepsilon$ and $\beta$ that span the excitability
plane shown in Fig.\ \ref{fig:parameterPlane}. Furthermore, the system is in
the state marked by the black full circle in Fig.\ \ref{fig:parameterPlane} and
the control goal is to reach a target state  being weakly susceptible to SD.
In this target  state, all SD waves with open ends shall retract or even
collapse so that SD occurs only transiently.  The whole regime of states being
weakly susceptible to SD  is located in the excitability plane to the left of
$\partial R_\infty$.  Between $\partial R_\infty$ and $\partial P$ only
retracting wave segments can occur (but see also Fig.~\ref{fig:sus} because the
location of the rotor boundary actually depends also on the size of  the wave
segment), and near but left to $\partial P$ in Fig.\ \ref{fig:parameterPlane}
only collapsing waves occur.  Then the seemingly simplest way to reduce tissue
excitability   is moving along arrow (A) in Fig.~\ref{fig:parameterPlane}. The
system then moves away from a location in the parameter regime of complex
re-entrant patterns revolving around a linear functional block
(Fig.~\ref{fig:parameterPlane}, inset), i.e., from a location to the right of
$\partial C$ (dotted line), towards a state in the regime being only weakly
susceptible to SD.

The efficiency of any control method that moves the system within the
excitability plane, for example along arrow (A)
(Fig.~\ref{fig:parameterPlane}), should be defined by some optimization
criterion. Path (A) is clearly better than  the path along the direction of the
arrow (B), because the latter path will not reach the target state, although
the excitability might seem to decrease moving along the direction of path (B),
if the dynamics of the functional core is used to probe excitability.  The core
motion becomes simpler to the left of $\partial C$ (dotted) and again simpler
crossing $\partial M$ (dashed): from complex to meandering to rigidly rotating
spiral-shaped SD waves, respectively (Sec.  \ref{sec:rd}).  This sequence of
patterns can erroneously indicate that eventually the system might become
weakly excitable as the same bifurcation lines need to be crossed before the
target state can be reached.  This generic behavior of complex systems poses a
problem, if the monitored functional core dynamics are used to probe the
effect of anti-migraine drugs. The same holds if even simpler wave parameters
such as the velocity are used  to probe the  effect of anti-migraine drugs
\cite{WIE96,BRA98b,SCH98r}. Moreover,  since a parameter plane has no metrical
structure, other measures than comparing paths that reach the target state to
those that do not are not available, in particular there is no privileged
shortest path. Without further constraints there exists an infinite number of
possible target states in the weakly susceptible regime. 


Let us now assume that excitability can only be controlled via nonlocal
transmission described by the operator $F$, in particular, via the influence of
long-range connections or time-delayed feedback. The control parameter is then
the coupling strength $K$ in Eq.\ \ref{eq:H}. We limit the nonlocal
transmission types to  the simplest coupling schemes with only one element of the
coupling matrix $A_{ij}$ being one and the others zero. This leads to four
coupling schemes, two self-coupling schemes with $i\!=\!j$ and two cross-coupling schemes with
$i\!\neq\!j$.  Other coupling schemes, for example diagonal coupling with
$A_{ij}\!=\!1$ for $i\!=\!j$ and zero otherwise,   are also possible and can be
treated in analogy, but the limitation to only one non-vanishing component $A_{ij}$ 
provides the four simplest algebraic constraints that allows us to evaluate
target states in the weakly susceptible regime of the excitability plane (Fig.\
\ref{fig:parameterPlane}), if the system is outside this regime.

Let the matrix element not equal to zero be denoted as $A_{ws}$ with $w,s \in \{u,v\}$,
so that the only augmented transmission is the following. The state from species $s$ is
used to create a nonlocal driving force $F[s]$ that is fed back to species $w$.

We first consider  the effect of  long-range connections on a planar SD wave
extending in the $y$-direction, so that only the propagation along the
$x$-direction needs to be considered. Moreover, we introduce the long-range
connections  only in one direction pointing against the direction of wave
propagation.  Without loss of generality, we let the wave propagate in the
negative $x$-direction, then this long-range connection leads to the additional term in
the rate equation of specie $w$ \begin{equation}
  \label{eq:anisotropic_backward_coupling}  F[s(x,t)] = s(x-\delta,t)-s(x,t).
\end{equation} where $\delta$ is the distance of the spatial long-range
connection.  This directed connectivity corresponds to anisotropic nonlocal
coupling in a two dimensional active medium. The functional and structural
connectivity of the cortex is realistically modeled as an anisotropic (and also
inhomogeneous) medium due to the patchy nature of nonlocal horizontal cortical
connections, however our  motivation to introduce Eq.\
(\ref{eq:anisotropic_backward_coupling})  was different. One reason is  to
obtain a better understanding of the results obtained in  previous work
\cite{SCH09c}, where isotropic nonlocal coupling have also been
considered.  Moreover, there is a direct analogy to local time-delayed
feedback.  In case of local time-delayed feedback the control force $F$ is
given by \begin{equation} \label{eq:TDAS} F[s(x,t)]= s(x,t-\tau)-s(x,t),
\end{equation} where $\tau$ is the delay time.

The common feature of the nonlocal transmission types in Eqs.\
(\ref{eq:anisotropic_backward_coupling})-(\ref{eq:TDAS}) is that the front of
the reaction-diffusion  wave described by Eqs.\
(\ref{eq:singleSpecies})-(\ref{eq:inhibitor_0}) receives an additional driving
force from the cortical steady state that lies ahead  of the wave (in space or
time).  The main idea, first introduced in Ref.\ \cite{SCH09c}, is to replace
the nonlocal terms, that is, the time-delayed term in Eq.~(\ref{eq:TDAS}), and
the space shifted term in Eq.~(\ref{eq:anisotropic_backward_coupling}),  by the
fixed point values $s(x,t-\tau)=s(x+\delta,t)=s^*$ of species $s$ for this is
the state  ahead of the propagating wave. For example, for the coupling
scheme $A_{vu}$ this special case derived from Eq.\ (\ref{eq:main}) with  Eqs.\
(\ref{eq:singleSpecies})-(\ref{eq:inhibitor_0}) becomes \begin{eqnarray}
  \label{eq:vu} \frac{\partial u}{\partial t} &=&  u - \frac{u^3}{3}\; - v
  \;+\;\nabla^2 u \\ \varepsilon^{-1} \frac{\partial v}{\partial t}&=& u +
  \beta - \gamma v + K (u^* - u),  \end{eqnarray} which can be transformed by
  simple algebraic operations 
\begin{eqnarray} 
    \label{eq:eepsilonvu}\tilde\varepsilon&=& (1-K)\varepsilon \\
    \label{eq:ebetavu}\tilde\beta&=&(\beta+K\,u^*)/(1-K)\\
    \label{eq:egammavu}\tilde\gamma&=&\gamma /(1-K)  
\end{eqnarray}
    to the reaction-diffusion system of activator-inhibitor type  Eqs.\
(\ref{eq:singleSpecies})-(\ref{eq:inhibitor_0}) with effective parameters
$\tilde \varepsilon$, $\tilde\beta$, and $\tilde\gamma$.

The effective parameters $\tilde \varepsilon(K)$, $\tilde\beta(K)$, and
$\tilde\gamma(K)$ in Eqs.\ (\ref{eq:eepsilonvu})-(\ref{eq:egammavu}) describe a
motion along a defined pathway parametrized by the coupling strength $K$ in
the parameter space of the extended Hodgkin-Grafstein model.   This pathway
will cross $\partial R_\infty$ at a certain point, while another coupling
scheme $A_{ws}$ can lead to a different pathway that crosses $\partial R_\infty$
at a different point for a different value of the coupling strength $K$. In
fact, it can be shown that the cross coupling scheme  $A_{uv}$ leads also to
Eqs.\ (\ref{eq:eepsilonvu})-(\ref{eq:egammavu}),   only with a change in the
sign of $K$ for the latter. Therefore, the cross coupling schemes $A_{vu}$ and
$A_{uv}$ do not differ with respect to their effect.  Note, however, that
whether they differ with respect to efficiency must be decided in a
dimensionalized model and there the ratio  $(\varepsilon A_{vu})/A_{uv}$ must
be used to compare the relative coupling strengths of the cross coupling
schemes, because the physical units must be the same (in analogy, the ratio
of the diffusion coefficients $D_v/D_u= \varepsilon \delta$, see Sec.\
\ref{sec:vt}).

By changing the coupling strength $K$  all three parameters are varied,
therefore we cannot directly illustrate the shift in the excitability plane
$(\varepsilon,\beta)$ in Fig.\ \ref{fig:parameterPlane}, which is a
cross-section of the full parameter space taken at $\gamma=0.5$.  How  the
effective parameters reduce susceptibility to wave propagations described by
either Eqs.\ (\ref{eq:eepsilonvu})-(\ref{eq:egammavu}) or equations
corresponding to the  other coupling schemes $A_{uu}$, $A_{uv}$, and $A_{vv}$,
was previously investigated in detail \cite{SCH09c}.   

The extended Hodgkin-Grafstein model is based on the Fitz\-Hugh-Na\-gumo
equation, so this approach describes   generic responses of reaction-diffusion
systems in the form of activator-inhibitor type to  control by nonlocal
coupling types given in Eqs.\
(\ref{eq:anisotropic_backward_coupling})-(\ref{eq:TDAS}).   We want to stress a
different perspective, namely that of a control failure and the subsequent
change of the cortical state that was before the failure in a state
nonsusceptible to SD.  An  internal nonlocal cortical control mechanism in the
form of $A_{ij}F$ that renders the cortical state nonsusceptible to SD and that
under pathological conditions is attenuated ($K \rightarrow 0$) can drive the
cortex into a state where retracting SD wave segments (weakly susceptible)
occur or even re-entrant SD waves.  This provides a putative common mechanism
of SD in migraine and stroke \cite{DAH08} whereas the cause of the changed
control pathway can be very different for the two conditions (cf. Fig.\
\ref{fig:sus} insets (a) and (b)).

\subsection{Stabilization of wave segments} \label{sec:stabilization}

\begin{figure}[t] \center \includegraphics[width=\columnwidth]{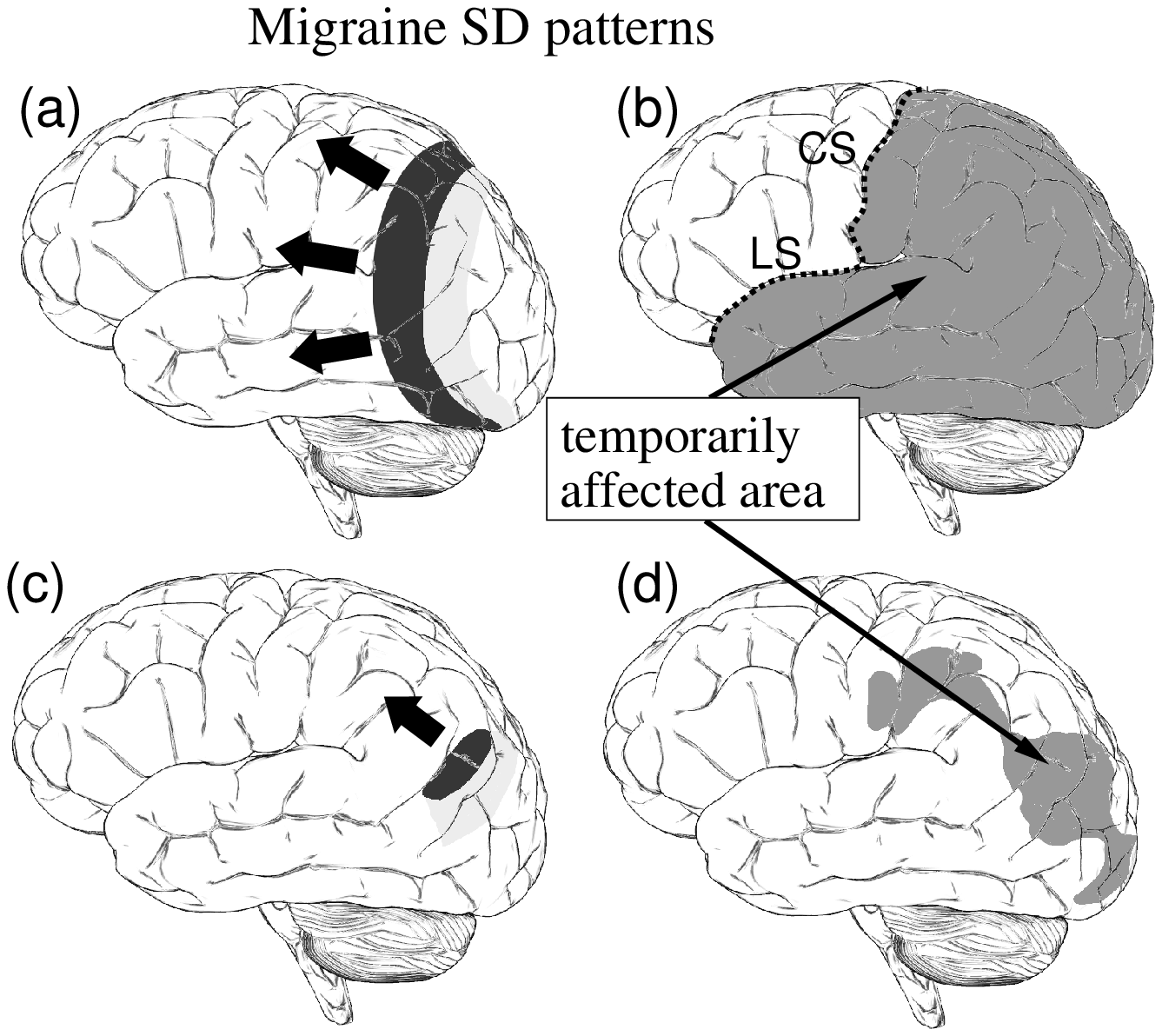}
  \caption{\label{fig:brains} Paradigms of SD propagation in migraine. Current
  view adopted from \cite{LAU87}: (a) SD wave engulfing a whole hemisphere.
  (b) A full-scale attack  affects most of the cortical surface (gray
  area). SD propagation is assumed to stop at the central and lateral sulcus
  (CS and LS, dotted line),  omitting the frontal lobe (not shaded).  Proposed
  change in paradigm: (c) localized SD wave segment affecting in a
  full-scale attack only a confined region of the cortex (b).  } \end{figure}

Fig.~\ref{fig:brains} (a) illustrates an SD wave that was triggered at some previous
time (about $10\,min$ earlier) in the rearmost portion of the human cortex (occipital
lobe).  Such figures became paradigmatic illustrations of SD wave propagation
in migraine with aura, found in modern textbooks of headache \cite{SIL05}. This
illustration made its first appearance  in a seminal paper spearheading the SD
theory of migraine more than 20 years ago \cite{LAU87}. It suggests that within
a single migraine aura attack lasting about $20\min$, half the cortex is
transiently invaded. Fig.~\ref{fig:brains} (b) shows  the spatial extent of a
"full-scale attack" \cite{LAU87}.  The wave stops in the middle, at the central
and lateral sulcus (CS and LS, dotted line), and omits the frontal lobe in
these illustrations. The sudden stop remains unexplained, although there are
speculations in the literature suggesting "some striking metabolic
difference of the two regions" \cite{REB67}.

The term {\em full-scale attack}  refers to the maximal spatial extension  of
an SD wave in migraine. In many attacks, the SD wave will stop before.  Even
taking into account less severe attacks, in which SD  stops earlier,  it has
been suggested that this paradigm  illustrated in Fig.~\ref{fig:brains} (a)-(b)
must  be modified, namely that the maximal extent of invaded cortical tissue
in a full-scale attack is far more limited (Fig.~\ref{fig:brains} (d)) caused
by a localized SD wave segment (Fig.~\ref{fig:brains} (c)) \cite{DAH08d}. The
instantaneously activated area  by SD in the human cortex is thus confined in
its width.  From a pattern formation point of view taken within the
framework of Eq.~(\ref{eq:main})-(\ref{eq:H}), this is  quite a different
picture revealing the presence of an additional nonlocal feedback mechanism
($K\ne0$).  In the next paragraph we put forward  arguments in favour of the
pattern in Fig.~\ref{fig:brains} (c)-(d). The last of these arguments hints at
the nature of a possibly involved nonlocal feedback represented by the operator $F$.


If  we assume that an SD wave engulfs on its course half the cortex as shown in
Fig.~\ref{fig:brains} (b), then only a subset of this activation results in
sensory, e.g., visual, awareness \cite{WIL04}. This seems rather unlikely as the
cellular depolarisation during  SD is the most dramatic cortical event known. Therefore,
the alternative is more likely, namely that SD in migraine occurs as a
localized wave segment that within a full-scale attack   affects cortical areas
along certain, though variable, paths (Fig.~\ref{fig:brains} (d)). This is
supported by a study investigating the sequence of neurological symptoms in
migraine \cite{VIN07}.  Moreover, similar propagation patterns were observed
for the gyrencephalic feline cortex \cite{JAM99}: while the primary SD wave
engulfed one cortical hemisphere, corresponding to Fig.~\ref{fig:brains} (b),
subsequent secondary waves remained within the originating gyrus, similar to
Fig.~\ref{fig:brains} (d) though more fragmented. Finally,  there is one study
using functional magnetic resonance imaging (fMRI) in humans during a migraine
attack \cite{HAD01}.  Although the spreading activity pattern observed by fMRI
seems to support the current paradigmatic view (Fig.~\ref{fig:brains} (a)), the
data is still inconclusive, as noted by Wilkinson \cite{WIL04}, because most of
the observed spreading activation could represent network activity through
feed-forward and feedback circuitry but not a depolarization wave. If this is
correct,  the global activity in the neural circuitry, while not being part of
SD, can modulate cortical susceptibility to SD. This kind of global activity
is, as we will show, exactly what is needed to understand stable
propagation of wave segments \cite{KRI94,SAK02,MIH02}.  

Suggesting that an SD wave in migraine takes the shape of a wave segment leads
to one main problem. Wave segments  are  usually not solutions to
reaction-diffusion systems of activator-inhibitor type, and therefore such
forms, chosen as an initial condition, transform quickly.  An open end of an SD
wave segment either curls in to form a spiral-shape wave
(Fig.~\ref{fig:parameterPlane} inset). This usually happens if cortical
susceptibility to SD is rather high. To be precise, open ends of sufficiently
large wave segments curl in, if the parameters $\varepsilon$ and $\beta$  of
the system lie towards the side of the rotor boundary $\partial R_\infty$ where
the value of the threshold  $\beta$ is lower and the value of the recovery rate
is slower (larger time scale ratio index $-\ln\varepsilon$) than it is on the
nearest point lying on $\partial R_\infty$. On the other side of the boundary
$\partial R_\infty$, wave segments independent of their size retract quickly.
The observed retinal SD wave patterns (Fig.~\ref{fig:rSDCocaine}) retract and
vanished within  a few millimeters (or, in temporal units, a few minutes).
This is in agreement with wave segments that are based on a reaction-diffusion
mechanism as described by
Eqs.~(\ref{eq:singleSpecies})-($\ref{eq:inhibitor_0}$) and simulated in the
weakly susceptible regime. They retract within the same order of magnitude
(note that for this estimation,
Eqs.~(\ref{eq:singleSpecies})-($\ref{eq:inhibitor_0}$) need to be
dimensionalized, see Ref.~\cite{DAH08}.) Similar results were obtained in a
descriptive mathematical model considering the motion of curves with free ends
\cite{DAH03a}.  In this geometric model,  spatially confined SD wave segments
were first suggested to cause aura symptoms, and it was estimated that a
full-scale migraine aura attack  hardly spreads more than several centimeters,
and this maximal distance is only obtained if parameters are correctly
adjusted.

\begin{figure}[t] \center
  \includegraphics[width=\columnwidth]{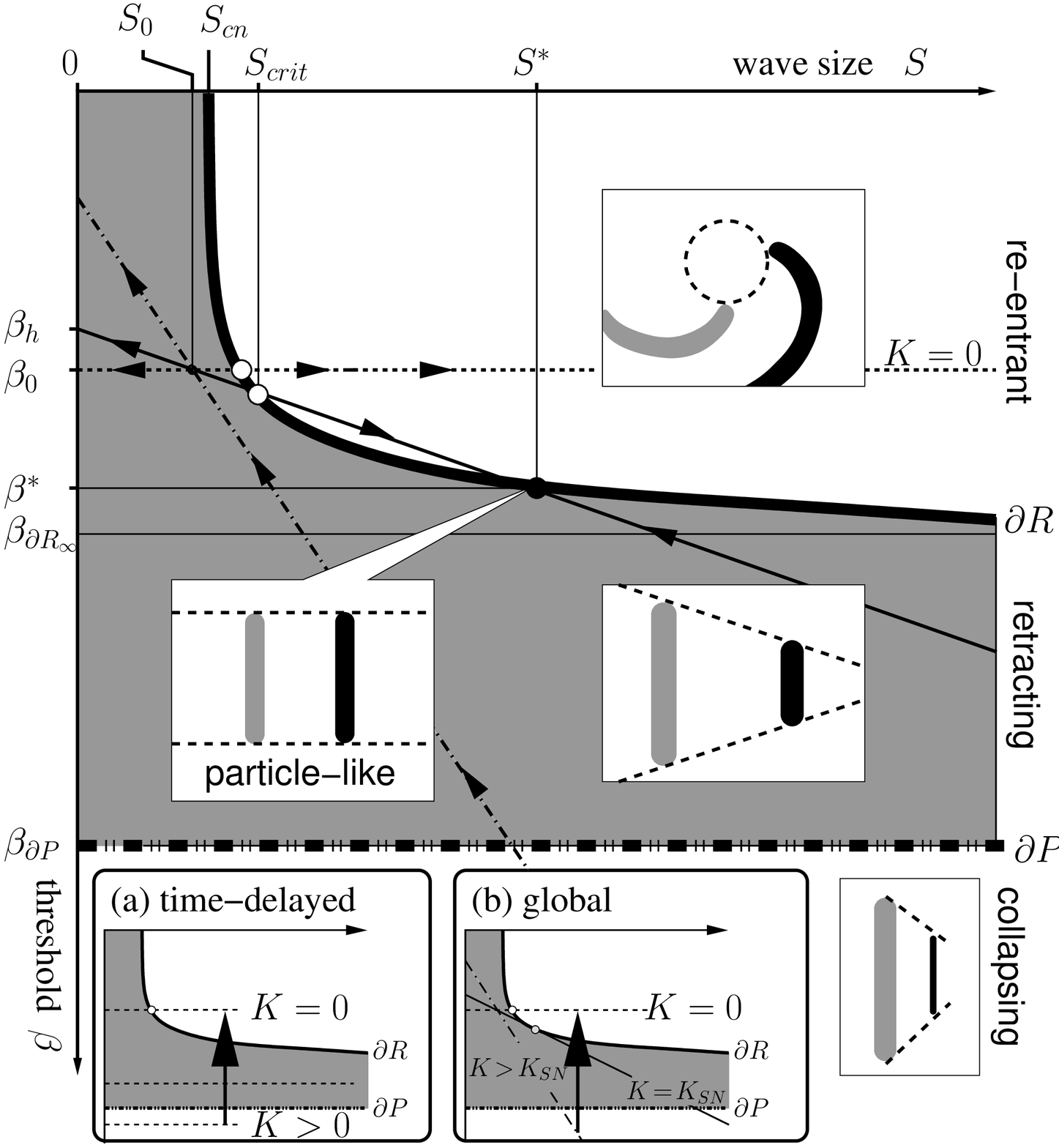}
  \caption{\label{fig:sus} Scheme of the classification of excitability
  according to spatio-temporal patterns (re-entrant/spiral, particle-like,
  retracting, collapsing) in 2D along the dashed-dotted line in
  Fig.~\ref{fig:parameterPlane}. The insets show wave front at two instances of
  time ($t_1$: gray, $t_2$: black) and trajectory of open wave ends (dashed).
  (a) The scheme illustrates a shift in the threshold by an attenuated ($K
  \rightarrow 0$) long-range coupling or time-delayed feedback according to
  Eq.\ \ref{eq:ebetavu}. (b) The  scheme illustrates a shear of the control
  line from dash-dotted to solid as global feedback (Eq.\
  \ref{eq:betaControlled}) is  attenuated ($K \rightarrow 0$). At $K=K_{sn}$
  (solid line) a saddle-node bifurcation occurs.} \end{figure} 

Adhering strictly to the definition of the rotor boundary as a boundary for
spiral waves, this boundary depends on the wave segment size $S$ and is called
$\partial R$.   At any point to the right  of $\partial R_\infty$ in Fig.\
\ref{fig:parameterPlane}    only SD wave segments above  a critical size curl
in at their two open ends  to form a counter-rotating double spiral wave, while
those below this size retract and the wave vanishes.  This can readily be
deduced from the fact that the two attracting final states coexist in this
parameter regime, namely the one with a counter-rotating double spiral wave and
the homogeneous steady  state. There is a separatrix manifold in the phase
space separating the two basins of these two attractors and thereby defining a
perturbation threshold.  Usually, on this separatrix manifold  a  fixed point
exists that is stable with respect to the flow parallel to this manifold.  This
fixed point is a saddle point in the full phase space and corresponds to an
unstable solution, a so-called particle-like wave of finite size  revealing a 
naturally  shaped wave segment that only retracts or grows when perturbed.
\cite{SAK02,MIH02}.  

The wave size $S$ can be defined   as the surface area of the wave segment
where  a certain elevated activator concentration $u_e$ is  exceeded
\begin{eqnarray} S(t) = \label{eq:waveSize} \int \Theta( u({\bf r},t) - u_{e})
  d{\bf r}, \end{eqnarray} where $\Theta$ is the Heaviside function and $\bf r$
  denotes the location on the two-dimensional cortical surface.  Since the wave
  size $S$ is a derived system quantity, it is not represented in the parameter
  plane in Fig.~\ref{fig:parameterPlane}, and therefore $\partial R$ cannot be
  represented either.  Instead,  the boundary $\partial R_\infty$ is marked
  as the limit of infinitely large wave segments  that neither retract nor
  grow, so-called critical fingers \cite{KAR91,HAK99}.  At $\partial
  R_\infty$, the cortical susceptibility to spiral SD waves   approaches its lower bound, because
  arbitrarily large segments retract below this boundary and therefore waves with open ends can not
  re-enter and will eventually disappear.

It was shown in a chemical model system of reaction-diffusion waves
experimentally and with a mathematical reaction-diffusion model
\cite{SAK02,MIH02} that particle-like waves can be stabilized  if the 
wave size $S(t)$, as given in Eq.\ (\ref{eq:waveSize}), is used to adjust
excitability $\phi$ by the linear relation \begin{eqnarray}
  \label{eq:globalControl}\phi(t) = a S(t) + b, \end{eqnarray} where $a$ and
  $b$ are characteristic feedback parameters of this global feedback control
  scheme.  In this chemical model system \cite{MIH02} the quantity $\phi$
  corresponds to the light intensity that controls the excitability of the
  chemical reactions, and this quantity  has a similar function as $\beta$ in
  the FitzHugh-Nagumo equations.

The global feedback control scheme in Eq.\ (\ref{eq:globalControl}) can be
translated into our framework by setting $a=K$ and $b=\beta_0 - K S_0$ with
some reference values $\beta_o$, $S_0$  and replacing in Eq.\
(\ref{eq:inhibitor_0}) the constant parameter $\beta$ with the controlled
threshold $\beta(t)=\phi(t)=a S(t) + b$ so that
\begin{eqnarray}\label{eq:betaControlled} \beta(t)=\beta_0 + K(S-S_0).
\end{eqnarray} For $K=0$, the uncontrolled FitzHugh-Nagumo inhibitor rate
equation with $\beta=\beta_0$  is obtained. The operator $F$ is then
\begin{eqnarray}\label{eq:globalF} F[u(x,t)] = S(t) - S_0 \end{eqnarray} and
  all components of the matrix fulfill $A_{ij}=0$ except for $A_{vu}=1$. $S_0$
  and $K$ are, like $a$ and $b$ in the notation of Eq.\
  (\ref{eq:globalControl}), the two characteristic feedback parameters of this
  global feedback control scheme.  In addition to these parameters, the value
  of $\beta_0$ plays a central role for this global control scheme. 

The meaning of $S_0$, $K$,  $\beta_0$, and further related quantities  can be
directly read from the plane in Fig.\ \ref{fig:sus}.  This  is an illustrating
scheme where the  vertical axis is spanned by the controlled threshold $\beta$
and corresponds in principle to the dash-dotted line marking a section in the
parameter space at constant $\varepsilon$ in Fig.\ \ref{fig:parameterPlane}.
The horizontal axis is spanned  by the size $S$ of the wave segment. Note that
we have dropped in the notation of $S$ and $\beta$ the explicit time
dependence, as this is more compact.  Throughout this section and in Fig.\
\ref{fig:sus}, $S$ and $\beta$ should be read as time dependent quantities.
Both quantities are projections form the infinite phase space of the
reaction-diffusion system, for $S$ this projection is defined via Eq.\
(\ref{eq:waveSize}) and for $\beta$ via  Eq.\ (\ref{eq:betaControlled}).  Since
$\beta$ is a quantity derived from  $S$,   only a one-dimensional section of the
$(S,\beta)$ plane corresponds to a phase space projection of a system with
parameters $S_0$, $K$, and  $\beta_0$, as described in the following.

Without the global coupling, i.\,e., for $K=0$, the value of the threshold
$\beta$ does not depend on $S$ and it is  therefore constant
($\beta=\beta_0$). In this case,  the evolution of wave segments represented in
the  $(S,\beta)$ plane is confined to a horizontal line, e.\,g., the thin
dotted horizontal line in Fig.\ \ref{fig:sus}. As described in previous Sec.\
\ref{sec:translation}, the value of the threshold of the free system ($K=0$)
can be in three characteristic regimes: (i) either it is in the regime where all
wave profiles collapse independent  of size and shape ($\beta_0>\beta_{\partial
P}$), (ii) or where all wave segments retract ($\beta_{\partial P} > \beta_0 >
\beta_{\partial R_{\infty}}$), (iii) or where only wave segments below a critical
size, marked by the solid thick black line $\partial R$, retract and if having
a larger size grow to form a spiral pattern ($\beta_0<\beta_{\partial
P_\infty}$).   Note that a horizontal line located at  $\beta=\beta_0$
indicates the absence of a global control scheme in Fig.\ \ref{fig:sus}, but
$\beta_0$ can still be an effective parameter as given in Eq.\
(\ref{eq:ebetavu}), that is,  $\beta_0$ can be, for example, being controlled
by time-delayed feedback or long-range coupling (inset (a)). 


With global coupling, i.\,e., for $K\neq 0$, the threshold  depends on $S$ and
the evolution of a wave segment represented in the  $(S,\beta)$ plane is then
confined to an 	inclined line, e.\,g., the thin solid and dash-dotted  lines
with arrows.  With increasing coupling strength $K$, the inclination with
respect to the horizon increases and the line of controlled wave segment
evolution  is sheared along the vertical line  $S=S_0$.  So there is one
invariant pivot  point $(S_0,\beta_0)$. If this point is located at certain
positions in the $(S,\beta)$ plane, and $\partial R$ is a convex boundary of
the re-entrant wave regime, there exists a value $K=K_{SN}$ at which the
inclined line becomes tangent to $\partial R$.  For values  $K>K_{SN}$
(dashed-dotted line, inset (b)), all wave segments retract   irrespectively of
their size, though the retraction will be slower than for any uncontrolled
system because the threshold is lowered as the wave segment size shrinks. At
$K=K_{SN}$, particle-like wave solutions are born in a saddle-node bifurcation.  

To obtain a saddle-node bifurcation at $K_{SN}$, the invariant pivot  point
$(S_0,\beta_0)$ must be located in either of two regimes: in the gray shaded
area below the value of $\beta_{\partial R_\infty}$, i.\,e., $S_0<\partial
P(\beta_0)$ and $\beta_0<\beta_{\partial R_\infty}$ (shown in Fig.\
\ref{fig:sus}), where $\partial P(\beta_0)$ denotes the projection of $S$ onto
$\partial P$ along $\beta_0$; or in the rectangular area defined by
$S_0>S_{cn}$ and $\beta_0>\beta_{\partial R_\infty}$.  If the invariant
pivot  point $(S_0,\beta_0)$ is located outside these regimes, again two
regimes should be distinguished.  If $(S_0,\beta_0)$ is in the rectangular area
defined by $S_0<S_{cn}$ and $\beta_0>\beta_{\partial R_\infty}$, there  exists
neither a value of $K$ for which the inclined line becomes tangent to $\partial
R$ nor will the inclined line of controlled wave segment evolution intersect
with $\partial R$ for $K>0$ and therefore particle waves cannot be stabilized.
If $(S_0,\beta_0)$ is in the white area below the value of $\beta_{\partial
R_\infty}$, i.\,e., $S_0>\partial P(\beta_0)$ and $\beta_0<\beta_{\partial
R_\infty}$ there is no critical value of $K_{SN}$ above which global control
successfully blows up the basin of attraction of the homogeneous steady state to
invade the whole phase space.

It is noteworthy to mention that  this control   adds a distinctly new
character to the spatio-temporal patterns by merely  changing the stability of
a solution that exists in the uncontrolled system. For example, for $K$
slightly smaller than $K_{SN}$, as indicated by the solid  line in Fig.\
\ref{fig:sus}, an initial wave segment above a critical wave size $S_{crit}$
will grow while the value of the controlled threshold $\beta$ will
simultaneously be adjusted by the global feedback to the final value $\beta^*$
at which a stable particle-like wave solution of size $S^*$ is obtained. This
particle-like wave is also a solution,  though unstable, of the
uncontrolled system Eqs.\ (\ref{eq:singleSpecies})-(\ref{eq:inhibitor_0}) with
$\beta_0=\beta^*$.  If, however, the initial wave segment size has a value below
the critical wave size $S_{crit}$, the wave segment will retract and once
vanished  the controlled threshold $\beta$ reaches the value $\beta_h$ for
the homogeneous state of the cortex. 

There are cases where $K$ is too large and  the control becomes  {\em too
hard}.  For example, for $K=0.1824$, $\gamma=0$, $\varepsilon=0.04$, and
$\beta_h=\beta_0-K S_0=1.721$, we observed that particle-like waves can start to
temporally {\em breathe}, that is, they  change their size in an oscillatory
fashion and reach the final target size $S^*$ only after a long transient of
several oscillations ($>50$).  From this we infer that the real parts of the
eigenvalues of the stable focus node, which corresponds to the stable
particle-like wave solution (full black circle), become close to zero.  A Hopf
bifurcation that leads  to stable breathing patterns, as observed in
semiconductor nanostructures \cite{SCH00,STE07,SCH09}, which are based on similar
dynamics, have, however, not been found.

It will be an important future task to relate concepts and quantities
illustrated in Fig.\ \ref{fig:sus} to SD in human cortex in order to develop
new therapeutic methods that make use of these concepts. That   Fig.\
\ref{fig:sus} correctly describes the control of cortical susceptibility to SD
in migraine is currently only supported by various correct predictions of such
a model.  Beside those mentioned in the outset of this paragraph and those
reported in previous studies \cite{DAH03a,DAH04a,DAH07a,DAH08d}, we note 
one additional prediction that shows the unifying character of this
framework. So-called clinically silent migraine aura, that is, an aura that
migraineurs are subjectively unaware of  because  they do not experience any
neurological disturbance, might simply be explained by a quickly retracting SD
wave patterns occurring for values  $K>K_{SN}$  (dash-dotted line). Likewise,
it is known that clinically silent epilepsies are caused by seizure activity that
does not break away from a focus. Our framework would still predict that
the control mechanism is switched on to suppress propagation, which  must result
in a blood flow 'fingerprint' of SD, which indeed is
observed by non-invasive imaging \cite{MOS08}.

\subsection{Re-entrant SD: a functional definition of tissue at risk}
\label{sec:anatomical}

\begin{figure}[t] \center
  \includegraphics[width=\columnwidth]{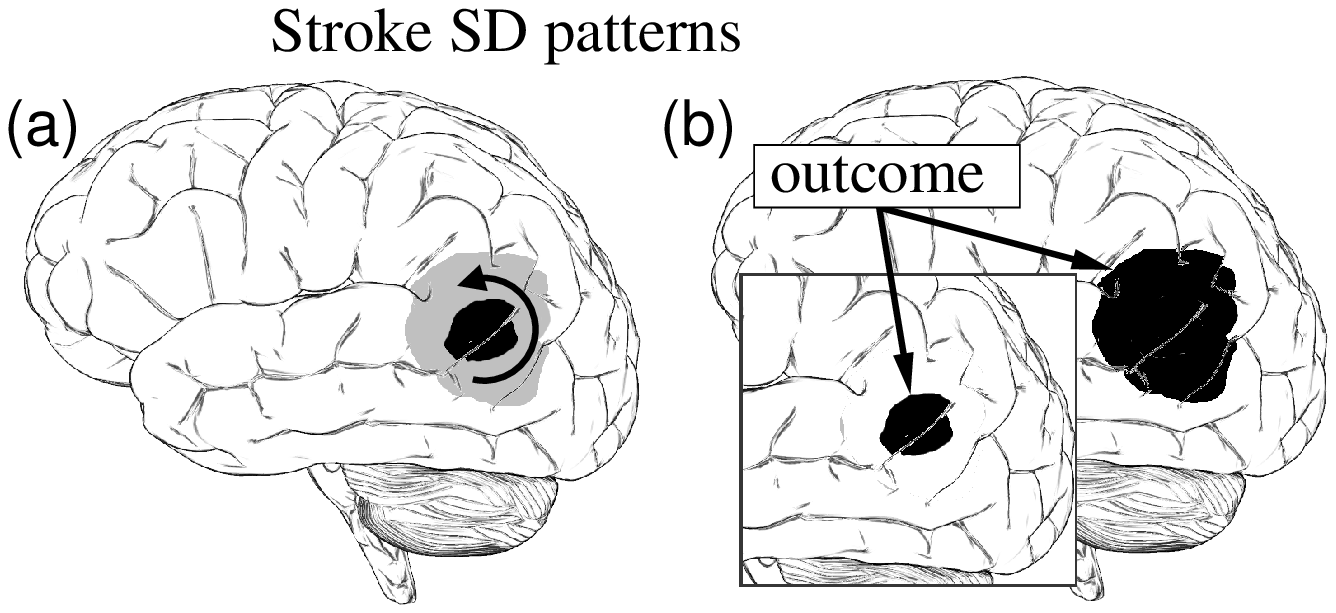}
  \caption{\label{fig:brainsStroke} (a) Cortical state during a localized
  graded reduction in blood flow.  Below a critical threshold of perfusion, an infarct lesion
  develops (black) surrounded by tissue becoming susceptible to SD waves
  (gray). These waves can spread in various patterns, e.g. radial outwards from
  the infarct core and cycling around it. (b) The infarct tissue may eventually
  include much of the surrounding tissue or be limited to the initial infarct
  core inset.  } \end{figure}

The re-entrant SD wave pattern observed in the experimental stroke model (Sec.\
\ref{sec:reentrantSD}) is illustrated in Fig.~\ref{fig:brainsStroke} (a): an
anatomical block of nonexcitable tissue (black) develops surrounded by a
ring-shaped zone becoming  susceptible to SD  (gray).  The experimental
procedure (Sec.~\ref{sec:cat}) does not necessarily cause   irrecoverable
damage in the form of an infarct lesion in the core region, but there is still
an anatomical block for a large central area is not affected by the
cycling. 

The focus in this section is on the surrounding ring-shaped zone (gray,
Fig.~\ref{fig:brainsStroke} (a)). Its dynamical features are of crucial
clinical interest, because  the aim of stroke therapy is salvage of this tissue
(Fig.~\ref{fig:brainsStroke} (b)).  At the outset of this section, we consider
the case that the anatomical block is an infarct core. In fact, when infarct
lesions  in the cerebral cortex are experimentally induced by vessel occlusion,
re-entrant patterns similar to those reported here emerge \cite{STR07}.   By
the end of this section, we will then reconsider that the anatomical block
induced with our experimental procedure is likely to be only nonsusceptible to
SD but also salvageable, in particular,  we compare the observed pattern with
strikingly similar observations in persistent migraine aura without infarction.

The cortical mechanisms by which the ring-shaped zone (gray,
Fig.~\ref{fig:brainsStroke} (a)) is formed are not yet fully established and
consequently there is also not a clear definition of how far this zone actually
spreads out. We suggest that the mathematical framework outlined in
Eq.~\ref{eq:main} can complement current clinical definitions of this spatial
structure.  The  zone surrounding an infarct core is either called penumbra
\cite{HOS06}, denoting a concept  based on  a decreased blood flow range, or an
alternative concept is based on the tissue at risk of infarction being
potentially salvageable. In stroke outcome, the infarct tissue may eventually
include much of the surrounding tissue (Fig.~\ref{fig:brainsStroke} (b)) or be
limited to the initial infarct core (Fig.~\ref{fig:brainsStroke} (b) inset).  

The {\em tissue-at-risk} (TAR) concept is clinically most relevant, but it does
not formulate a distinctive definition.  This concept must  be substantiated by
what processes underlie this risk so that it becomes clear what constitutes
high and low-risk.  A  decreased  blood flow range is one possible
substantiation, which, if the sole complement,  would identify the penumbra as
TAR.  The penumbra is anatomically defined. By contrast TAR can also be defined
as a functional concept in terms of nonlinear dynamics \cite{DAH07a}.  In its
original functional form, TAR was suggested to extend over the area invaded by
transient waves occurring below the propagation boundary (to the right of
$\partial P$ (Fig.~\ref{fig:parameterPlane}). This functional  definition did
not incorporate a gradient in susceptibility caused by anatomical
heterogeneities.  We suggest in the following to combine these  anatomical  and
functional concepts, which would then attach great clinical´ importance to
re-entrant pattern.

The defining property of re-entrant pattern with an anatomical block is the
heterogeneity.  In stroke, this heterogeneity is evolving and occurs actually
over several scales.  On the largest scale is the infarct core and the
surrounding region. In the infarct core blood supply is below the threshold of
energy failure, outside it is constrained but the energy state is preserved.
On a smaller scale occurs a heterogeneous microcirculation  \cite{LEN77,
HOS06}.  A mathematical model that does not incorporate a complete
characterization of this heterogeneous microcirculation can use nonlocal terms
to describe the macroscopic spatial connectivity in form of a  network of local
excitable elements described by
Eqs.~(\ref{eq:singleSpecies}-(\ref{eq:inhibitor_0})).  In such a model
\cite{GLA06,GLA07}, it was shown that spatial  diversity  induces wave
patterns.  Diversity is able to induce a transition to an excitable behavior of
the net.  These patterns are most coherent for an intermediate variability
strength, an effect that is described as similar to stochastic resonance
generated by additive noise.

\begin{figure}[t] \center \includegraphics[width=\columnwidth]{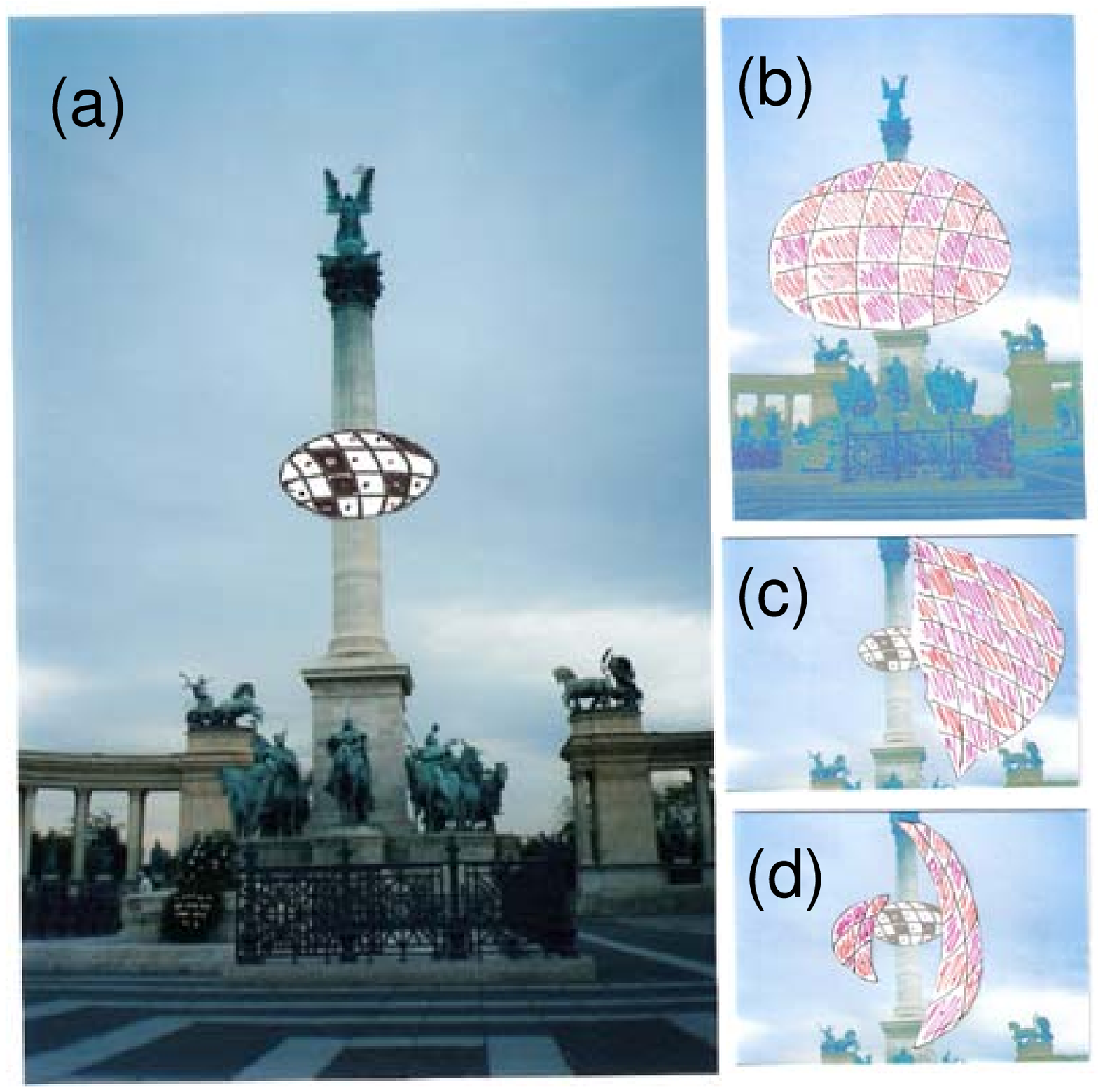}
  \caption{\label{fig:persistent} (a) Persistent central scotoma seen with open
  eyes. The area of the central scotoma is filled by geometric hallucinations
  of lattice form dimension. The pattern in this area remains when eyes are
  closed. During a migraine attack starting from the central scotoma various
  patterns can occur. (b) Expanding the initial affected area in the visual
  field, (c) waves propagation in one visual hemifield, and (d) waves
  propagating in both visual hemifields.}  \end{figure}

The spatio-temporal pattern reported  in Sec.\ \ref{sec:reentrantSD} are
strikingly reminiscent of patterns reported by migraineurs suffering from a
rare but well documented subform of migraine classified as {\em persistent
migraine aura without infarction}  in the  second edition of the international
headache classification (ICHD-II code 1.5.3).  In this subform, migraine aura
phenomena, which are otherwise transient neurological symptoms lasting less
then $30\,min$, persist for more than 1 week, maybe months or years.  To
diagnose this subform, there must not be any evidence by noninvasive imaging of
an infarct and these symptoms must not be attributed to another disorder.
In Fig.\ \ref{fig:persistent} (a) a persistent central visual disturbance
is illustrated. The scotoma is filled by geometric pseudohallucinations of
lattice form.  While this pattern is seen always, even with eyes closed during
an acute migraine attack the  size of central scotoma can decrease and  become
up to 4 times larger (Fig.\ \ref{fig:persistent} (b)), or a further disturbance
can propagate in one visual hemifield (c) or in both (d).

In analogy to the stabilization of particle-like waves described in Sec.\
\ref{sec:stabilization}, these stationary patterns might be explained by the
stabilization of a critical nucleus with global feedback. The critical nucleus
refers to a spatially confined structure lying in the  phase space on the
separatrix manifold separating  the basins of attraction of a ring-shaped wave
and the homogeneous steady state.  It defines a stationary radial symmetric
perturbation threshold, while the particle-like wave is a traveling asymmetric
perturbation threshold.  The stabilization of the critical nucleus can probably
not be achieved by global feedback as introduced in Eq.\ (\ref{eq:globalF}),
because the coupling strength must be very large.  A large inclination of the
control line in  Fig.\ \ref{fig:sus} can be predicted if assuming that the
particle-like wave size approaches the size of the critical nucleus $S_{cn}$
for small thresholds.  However, if the operator $F$  has a nonlinear dependence
on $S$, or additional long-range, time-delayed, and other nonlocal augmented
transmission schemes, including spatial  diversity, are considered in
combination with reaction-diffusion, a rich repertoire opens up and it will
be an important task for future research to investigate this in the context of
migraine and stroke patterns.

\section{Discussion and outlook}
\label{sec:discussion}

We have presented a mathematical framework for cortical spreading depolarizations
that unifies existing activator-inhibitor models as special cases
\cite{GRA63,TUC81,REG94,REG96,DAH04b}. Within this framework the
reaction-diffusion mechanism of SD is extended by nonlocal coupling as an
integral part of the phenomenon necessary to describe the emergence and
transitions of two-dimensional SD wave patterns. As a framework it provides a
flexible procedure for analysing SD patterns observed in both experimental
migraine and stroke models, and clinical data \cite{HAD01,DOH08}.  

This framework is aimed at providing basic insights needed to understand
cortical susceptibility to transient SD wave propagation in terms of nonlinear
science. In particular, we focus on bifurcations in the dynamics of cortical
homeostasis  caused by cortical circuits and neurovascular coupling represented
as nonlocal transmission schemes.  A long term goal is to provide control
strategies based on bifurcation theory for both traditional pharmacological
treatments and biomedical engineered devices that intelligently target the
occurrence of SD waves. 

A mathematical model that involves only two species with activator-inhibitor
dynamics  is undoubtedly a very crude macroscopic level of cortical
homeostasis. Activator-inhibitor dynamics is the minimal requirement that leads
to a description of SD as a traveling wave pattern with a trailing edge. In two
spatial dimensions, these waves take a characteristic  form that reveals much
about cortical susceptibility to SD.  Yet, little insight into microscopic
mechanisms of SD is offered by such minimal dynamic systems.  There are much
more sophisticated mathematical descriptions of SD on a detailed microscopic
level that try to resolve the mechanism of SD \cite{KAG00,MAK07}. In our framework, these
microscopic mechanism   are hidden inside the box labeled as "firing rate"
(Fig.~\ref{fig:sdPathways}).

Proposing a conceptual framework at this macroscopic level must be justified by
showing that this approach avoids deficiencies that would arise if SD wave
propagation is modeled on a microscopic level.  From the clinical perspective,
the macroscopic scale, in particular, the two-dimensional pattern on the
cortical surface,  is  highly relevant. In migraine, the whole spectrum of aura
subtypes could depend critically on whether SD can break away from a focus
point,  and, if it does, whether SD waves take the shape of a
particle-like wave and therefore limit the extent of a full-scale attack to
certain cortical regions (Fig.\ \ref{fig:brains} (d)), or, in the worst case, SD
becomes a persistent activity pattern (Fig.\ \ref{fig:persistent}). In stroke,
outcome depends on SD as these waves are suspected to contribute to the loss of
potentially salvageable tissue, i.\,e., tissue at risk of infarction. Since
re-entrant pattern increase the total number of waves, the emergence of these
patterns need to be understood.  None of these questions can be addressed in a
cellular model of SD including only one or several neurons,  let alone the
semantic objection against applying the term {\it spreading}  to such models
because they are not meant to model the spread \cite{SOM01}.

Microscopic models include several ion channel types in a single neuron, and
its dendritic tree. These models also consider  surrounding compartments, and
can therefore also describe changes in intra- and extracelluar ion
concentrations \cite{KAG00,MAK07}.  To obtain the continuum limit of such
discrete microscopic cellular models is by no means straightforward.  The
problem is to match the very detailed and accurate  knowledge of microscopic
processes involved in the depolarization cycle of SD by an equally detailed
description of volume transmission and wiring transmission needed to provide
the spatial coupling for transition to the continuum limit. Mismatching these
levels of description by merely  using a diffusion term as in Eq.\ (\ref{eq:D})
to simulate spatial coupling by extracellular potassium would be like attaching
the wheels of a carriage on a Mercedes.

Efforts have been made to address this problem by accounting not only for
interstitial ionic diffusion but also for ionic movement through a neuronal
syncytium of cells connected by gap junctions  and for cells that are allowed
to expand in response to osmotic pressure. From these it was concluded that
cytosolic diffusion via gap junctions and osmotic forces are important
mechanisms underlying SD \cite{SHA01}. Yet, to compute the spatio-temporal
development of SD across the cortical surface on a centimeter scale and
investigate the emergence of retracting, re-entrant, and stationary SD waves,
coupling schemes other than local schemes must be accounted for. There are
important nonlocal transmission schemes in the cortex hat provide spatial
coupling over a long range or lead to time delays. Example are (i) the
functional and structural cortical connectivity, in the continuum limit
described by neural fields \cite{WIL73}, (ii) the cortical energy state
maintained by a segmental blood perfusion from the arteries to smaller
arterioles and finally branching capillaries leading to a complex
hemodynamic response that is regulated   by smooth muscle cells (arterioles)
and pericytes (capillaries) \cite{PEP06}, and (iii) the topology of the network
of the neuronal syncytium that will be compromised by disseminate neuronal
injury during ischemia \cite{HAN97a}.  There are  more transmission schemes,
but the effect of these (i)-(iii) are the ones we have considered in a first
approximation by (i) long-range connections, (ii) global coupling, and (iii)
spatial  diversity.

Future investigations have to relate the quantitative values of the parameters
in macroscopic model that successfully describe the bifurcations observed in
the spatio-temporal characteristics of SD to the cellular level. This will open
up this approach for the design of both drug treatments \cite{DAH07a} and
engineered devices utilizing transcranial stimulation \cite{LIE06} with the aim
to  target SD intelligently.  Migraine and stroke are found in specific genetic
disorders giving clues to genetic factors that hint at differences in the
cortical network activity. Such hints may provide means to bridge the gap
between the macroscopic and microscopic level of SD.  Similar hints of changed
network activity are coming from psychophysics describing abnormal cortical
processing in migraine by concepts like  hyper- and hypoexcitability,
heightened responsiveness, a lack of habituation and/or a lack of
intra-cortical inhibition \cite{SHE01,WIL04}. Our framework provides means by
which such statements on cortical excitability in migraine  can be
investigated, in particular  how  abnormal cortical nonlocal connectivity
changes susceptibility to SD.  Although it seems tempting to suggest that
cortical hyperexcitability increases susceptibility to SD or even that neurons
prone to hyperexcitabilty trigger SD, such a simple relation cannot be
expected, and a detailed bifurcation analysis will be important for the
thorough understanding of SD as a nonlinear pattern formation process.






\section*{Acknowledgement} We would like to thank Steve Coombes, Felix
Schneider, Oscar Herreras, Hannelore Rittmann-Frank, Henry Tuckwell, Anthony
Gardner-Medwin, and Wyste Wadman for fruitful discussions at the Nottingham
migraine meeting, and Vladimir Zykov for comments on particle-like waves.
MAD was supported by the Deutsche Forschungsgemeinschaft (DA 602/1-1 and SFB
555)

\end{document}